\PassOptionsToPackage{dvipsnames}{xcolor}
\documentclass[preprint]{aastex631}

\pdfoutput=1

\usepackage{amsmath}
\usepackage{graphicx}

\newcommand\eg{\textit{e.g.}}

\usepackage[normalem]{ulem}
\usepackage{CJK}

\shorttitle{Stellar Flare Late Phase}
\shortauthors{Yang et al.}
\graphicspath{{./}{figures/}}

\begin{document}
\begin{CJK*}{UTF8}{gbsn}

\title{A Possible Mechanism for ``Late Phase'' in Stellar White-Light Flares}

\author[0000-0002-7663-7652]{Kai E. Yang (杨凯)}
\author[0000-0003-4043-616X]{Xudong Sun (孙旭东)}
\affiliation{Institute for Astronomy, University of Hawai`i at M\=anoa, Pukalani, HI 96768, USA}

\author[0000-0001-5316-914X]{Graham S. Kerr}
\affiliation{NASA Goddard Space Flight Center, Heliophysics Science Division, Code 671, Greenbelt, MD 20771, USA}
\affiliation{Department of Physics, Catholic University of America, Washington DC 20064, USA}

\author[0000-0001-5685-1283]{Hugh S. Hudson}
\affiliation{SUPA School of Physics and Astronomy, University of Glasgow, Glasgow G12 8QQ, UK}
\affiliation{Space Sciences Laboratory, University of California, Berkeley, CA 94720, USA}

\correspondingauthor{Kai E. Yang}
\email{yangkai@hawaii.edu}

\begin{abstract}
M-dwarf flares observed by the \textit{Transiting Exoplanet Survey Satellite} (\textit{TESS}) sometimes exhibit a ``peak-bump'' light-curve morphology, characterized by a secondary, gradual peak well after the main, impulsive peak. A similar ``late phase'' is frequently detected in solar flares observed in the extreme-ultraviolet from longer hot coronal loops distinct from the impulsive flare structures. White-light emission has also been observed in off-limb solar flare loops. Here, we perform a suite of one-dimensional hydrodynamic loop simulations for M-dwarf flares inspired by these solar examples. Our results suggest that coronal plasma condensation following impulsive flare heating can yield high electron number density in the loop, allowing it to contribute significantly to the optical light curves via free-bound and free-free emission mechanisms. Our simulation results qualitatively agree with \textit{TESS} observations: the longer evolutionary time scale of coronal loops produces a distinct, secondary emission peak; its intensity increases with the injected flare energy. We argue that coronal plasma condensation is a possible mechanism for the \textit{TESS} late-phase flares.
\end{abstract}

\keywords{Stellar flares (1603); Solar flares (1496); Solar coronal loops (1485); Hydrodynamical simulations (767)}

\section{Introduction}\label{sec:intro}

The number of the observed stellar flares has increased dramatically since the launch of the \textit{Kepler} mission \citep{Maehara2012,Balona2015,Davenport2016,VanDoorsselaere2017,Yang2019,Notsu2019} and the \textit{Transiting Exoplanet Survey Satellite} (\textit{TESS}, \citealt{Gunther2020,Tu2020,Tu2021,Crowley2022,2022Pietras}). Both stellar and solar flares are thought to be driven by the release of magnetic energy from magnetic reconnection \citep{2002A&ARv..10..313P,Benz2010}, and while they share many common characteristics stellar flares can be significantly more energetic than their solar cousins. 
 Stellar superflares can release up to $10^{33}$--$10^{36}$ erg bolometric energy \citep[e.g.][]{Maehara2012,Gunther2020} compared to solar flares, the greatest of which release less than or equal to about $10^{32}-10^{33}$ erg \citep[see discussions of the Carrington flare level in][and references therein]{2023Hayakawa}. This makes space-weather conditions around these stars more hazardous than in the heliosphere.
It is worth noting that some of the stellar flare have also been recorded with energy below $10^{30}$ erg \citep{Howard2022,2022Pietras}.

The majority of optical stellar flare light curves exhibit an exponential rise and a gradual decay \citep[\eg][]{Davenport2014}. Such an impulsive peak is thought to originate from the low stellar atmosphere \citep[\eg][]{hudson2006}.
A fraction of flare lightcurves also display finer features such as quasi-periodic pulsations \citep[QPP;][]{Rodono1974,Zimovets2021}.
Based on certain solar observations and models, studies have suggested that some QPPs may result from the dynamic response of the stellar atmosphere to impulsive heating \citep{VanDoorsselaere2016,Nakariakov2009,Zimovets2021}.

\begin{figure}[t!]
\centering
\fig{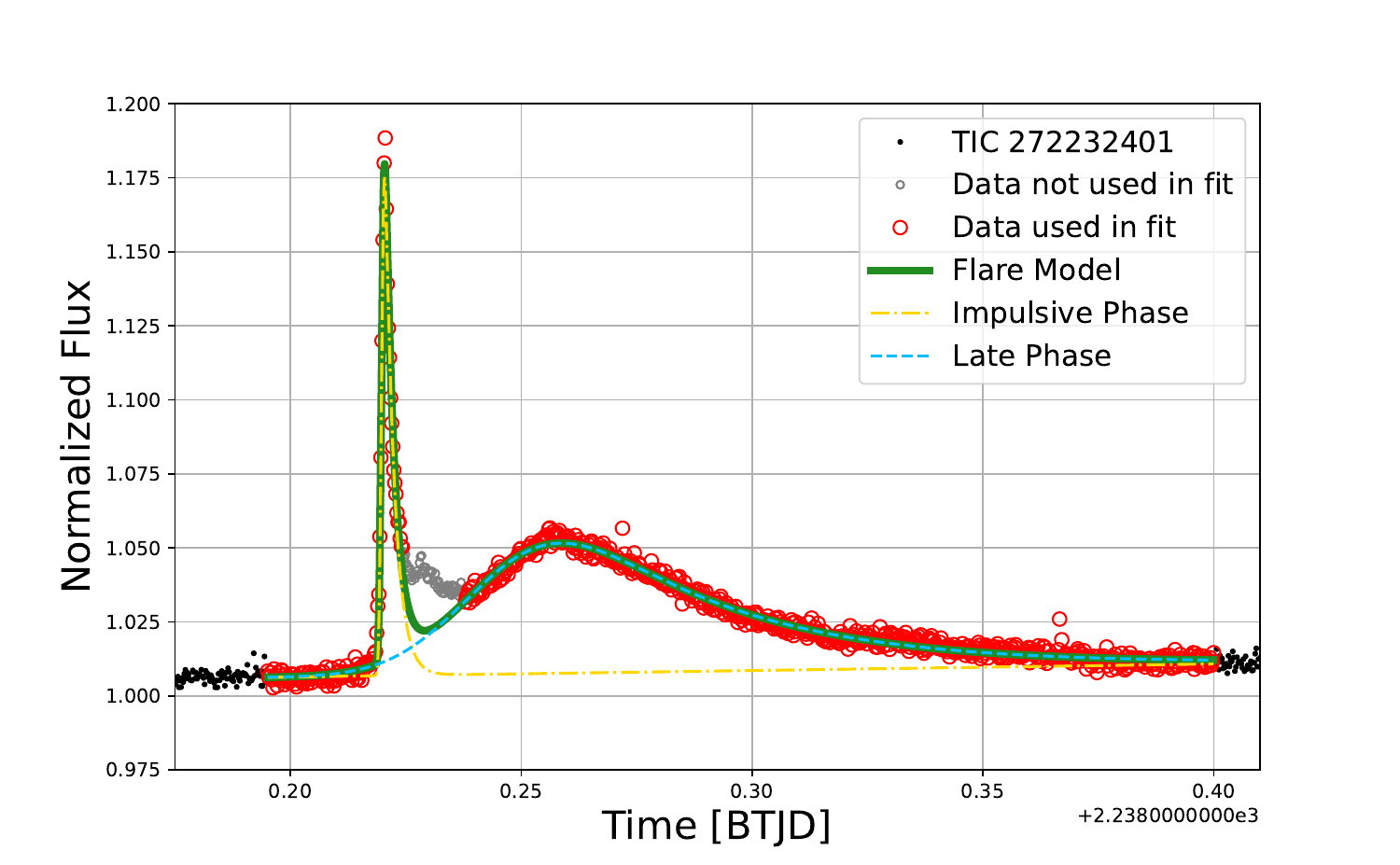}{1\textwidth}{}
\caption{An example of \textit{TESS} flare light curve with a ``peak-bump" morphology from the \textit{TESS} Input Catalog (TIC) ID 272232401. The x-axis is a fraction of a day starting from Barycentric \textit{TESS} Julian Date (BTJD) 2238. Black dots represent data outside the fitting range, red circles indicate data used for fitting, and gray circles denote excluded data around time $2238.23$. These data are excluded due to a small local maximum that may come from another region of the star and is not well described by the ``peak-bump" morphology. The light curve is fitted using an empirical ``double flare'' template $f(t)$ from \cite{2022Pietras}: $f(t)=\int_0^t (A_1\exp[-(x-B_1)^2/C_1^2]\exp[-D_1(t-x)]+A_2\exp[-(x-B_2)^2/C_2^2]\exp[-D_2(t-x)])\mathrm{d}x$, where $t$ represents time. The fitted impulsive phase (first portion of $f(t)$), late phase (second portion of $f(t)$), and total flare model are shown as gold dash-dot, cyan dashed, and green solid lines, respectively. The best-fit parameters are: $A_1=8.52$ hr$^{-1}$, $B_1=1.07$ hr, $C_1=1.47\times10^{-2}$ hr, $D_1=2.11\times10^1$ hr$^{-1}$, $A_2=8.3\times10^{-2}$ hr$^{-1}$, $B_2=1.6$ hr, $C_2=4.96\times10^{-1}$ hr, $D_2=1.12$ hr$^{-1}$. The $\chi^2 $ of the fit is $2.72$.
\label{fig:1}}
\end{figure}

In a recent survey, \citet{Howard2022} found $42.3\%$ of $3792$ M-dwarf ``super flares'' (energy above $10^{32}$~erg) observed by \textit{TESS} showed complex morphology in their light curves \citep{Howard2022}. Furthermore, $17\%$ (31 total) of these complex flares exhibited a ``peak-bump" morphology, with ``a large, highly impulsive peak followed by a second, more gradual Gaussian peak.'' In another survey of $\sim$140,000 flares from over 25,000 late-type stars, \citealt{2022Pietras} found that approximately $40\%$ of intense long-duration flares can be described by ``double flare'' profiles, with the second component dominating the decay phase. As shown in Figure \ref{fig:1}, we choose one ``peak-bump" flare from \cite{Howard2022} as an example, and fit the light curve using a double-flare profile \citep[see Eq. 3 in][]{2022Pietras}.
The secondary, ``late phase'' has a distinct time scale of tens of minutes, and a significant amplitude, typically a few percent of the stellar flux. Its occurrence rate was too high to be explained by chance occurrence.

The peak-bump morphology is in fact frequently detected in solar flares, typically in the extreme ultraviolet (EUV) and soft X-ray (SXR) wavelengths \citep{Woods2011,Qiu2012,Liu2013,Liu2015,Sun2013,Dai2013,Dai2018a,Dai2018b,Chen2020}.
The EUV/SXR late phase originates from long coronal loops that are magnetically connected to the main flare site. During the impulsive phase, these loops are rapidly filled with high-temperature plasma, which requires longer time to cool via conduction compared to shorter loops \citep{Reale2014}. A secondary peak will be delayed in the spectral irradiance of cooler lines, with a time scale ranging from tens of minutes to hours. Additional heating and further magnetic reconnection may also contribute to the delay, and produce late phase `hot' emission at somewhat lower temperatures than in the primary event.

What is the cause of the late phase in \textit{TESS} optical light curves? Does it originate from the coronal loops, similar to the solar counterpart observed in hotter EUV/SXR emitting plasma? Before answering these questions, we note that the solar flare energy content is on the lower end of the known stellar flare population. Its impact on total solar irradiance is at most a few $10^{-5}$ \citep{Kretzschmar2010}, orders of magnitude less than the detected stellar counterparts. Spatially resolved solar white light flare observations suggest that the optical continuum enhancements originate from the lower atmosphere\footnote{It is not clear, yet, if the continuum originates from the deepest, optically thick, layers of the photosphere/upper photosphere (appearing as an enhanced blackbody-like spectrum) or from higher altitudes in the mid-upper chromosphere (resulting from overionization and the subsequent enhanced recombination spectrum). Evidence exists for both origins \citep[e.g., see discussions in][and references therein]{2014ApJ...783...98K,2016ApJ...816...88K}.}. The locally integrated light curves are mostly single peaked, with rare exceptions \citep[e.g.,][]{Matthews2003,2014ApJ...783...98K,Hao2017}.

Off-limb white-light structures have been observed in the gradual phase of some solar flares, implying a significant density enhancement in the coronal loops \citep[e.g.,][and references therein]{1992PASJ...44...55H,Fremstad2023}. More examples of such observations \citep{MartinezOliveros2014,Fremstad2023} have been obtained from the Helioseismic and Magnetic Imager \citep[HMI;][]{Schou2012}, which provides pseudo-continuum images derived from the \ion{Fe}{1} 6173.3~\AA\ line observations.
The most famous recent example is from the bright solar flare \texttt{SOL2017-09-10T15:35}. Situated on the west limb, it hosted post-reconnection coronal loops emitting brilliantly in the continuum against the background \citep{Jejcic2018, Zhao2021, MartinezOliveros2022, Fleishman2022}. The loop apexes reached about $18$~Mm ($25\arcsec$); the continuum emission in those loops peaked about 30 minutes after SXR peak.

Based on such observations, \cite{Heinzel2017} and \cite{Heinzel2018} proposed that optically thin emission from the corona can meaningfully contribute to the white-light stellar flare emission. 
Their calculations showed that the intensity of the observations can constrain the electron number density ($n_\mathrm{e}$) and temperature ($T$) within a specific parameter regime. A high electron density $n_{e}\gtrsim10^{12}$--$10^{13}$~cm$^{-3}$ is unequivocally required, leading to free-free and free-bound emission mechanisms dominating over Thomson scattering. While such high coronal densities have indeed been reported for some solar flares \citep{Hiei1992,Heinzel2017,Jejcic2018}, 
it remains unclear what physical mechanism can generate the specific $(n_\mathrm{e}, T)$ values sufficient to produce intense continuum enhancements in the coronal loops \cite[see, for example, discussion in Section 2.3 of][]{Kerr2023}.

We note that another solar phenomenon, known as ``coronal rain'' \citep[e.g.,][and references therein]{Scullion2016,Mason2022,Antolin2022}, results from the rapid increase in local coronal density by orders of magnitude after the flare impulsive phase.
It is generally explained by thermal instability \citep{Parker1953,Field1965,Chaes2019,Claes2020}, where the plasma cooling rate increases drastically due to the density enhancement related to radiative energy loss. 
In a detailed magnetohydrodynamic (MHD) simulation, condensed plasma in the form of coronal rain appeared tens of minutes after the onset of magnetic reconnection \citep{Ruan2021}.
This mechanism has been evoked to explain the \textsl{in-situ} formation of solar prominences \citep{Xia2016,Kaneko2017,Zhou2020}, and can potentially produce dense recombining plasma capable of white-light flare emission.

In this work, we propose coronal plasma condensation\footnote{Note that this is distinct from ``chromospheric condensations'' that are often discussed in the context of solar flares. Those are dense downflowing regions in the chromosphere resulting from energy deposition in the impulsive phase. For the remainder of this manuscript, if we say ``condensation'' we are referring to coronal condensations. The term ``condensation" refers to the cooling process experienced by the plasma in the solar corona due to thermal instability. This cooling leads to a decrease in pressure and the subsequent formation of highly compressed plasma with an increased density in several orders, which is commonly used in recent literature \citep{Antolin2022}. Our usage of this term differs from its historical meaning, which was introduced by \cite{Waldmeier1963} to describe the forbidden-line corona.} as a possible mechanism for the observed ``peak-bump'' morphology in \textit{TESS} stellar flare light curves. 
To probe the underlying mechanism, we perform a suite of one-dimensional (1D) hydrodynamic (HD) loop simulations for typical M-dwarf flare parameters. We show that the simulated, optically thin white-light emission, in terms of the evolution time scale and the total flux, are qualitatively in agreement with \textit{TESS} observations. Below, Section~\ref{sec:model} describes our 1D HD simulations and forward synthesis of the \textit{TESS} light curve. Section~\ref{sec:result} presents our modeling results. Section \ref{sec:discussion} presents a summary and discussions.

\section{Model}\label{sec:model}

\subsection{HD Simulation}\label{sec:sim}

The solar flare occurs in a 3D volume, but performing parameter studies efficiently and simulating a realistic chromosphere exceeds current computational capabilities of 3D models. 
However, solar flares consistently occur in loop structures that follow the geometry of magnetic field lines in the corona \citep{Benz2017}.
These magnetic field lines guide the flow of hot plasma and produce coronal loops with enhanced emission.
The standard solar flare model suggests that magnetic reconnection heats the plasma in the loop, converting magnetic energy into heat, bulk motions, and radiation emissions such as X-rays. 
Due to the confined and elongated nature of the loop, it can be approximated as a 1D structure for modeling purposes, which simplifies and facilitates simulations using 1D hydrodynamic models \citep[\eg][]{McClymont1983}.
This approach has been successfully used in many solar flare simulations, including the physics of energy and radiation transport, such as nonthermal particles, Alfv\'en waves, thermal conduction, and radiation \citep[][and references therein]{Kerr2022,Kerr2023}. 
One-dimensional models have also been employed to simulate M-dwarf flares (\citealt{Allred2015,Kowalski2023}).

We use the open-source Message Passing Interface Adaptive Mesh Refinement Versatile Advection Code 2.0 \citep[MPI-AMRVAC,][]{Xia2018, Keppens2021} to simulate a flaring loop via a 1D hydrodynamic model.
The MPI-AMRVAC code is a parallized partial differential equation solver framework that contains many different numerical schemes in multi-dimension with multiple physics modules. This code has been successfully applied to simulating solar flare/eruption \citep{Fang2016,Ruan2020,Ruan2023,Zhong2021,Zhong2023,Guo2021,Guo2023,Druett2023} as well as the formation and evolution of cold material in the solar corona \citep{Xia2016,Xia2017,Zhou2018,Zhou2020,Zhou2023,Hermans2021,Jenkins2022}.
For the simulation in this paper, we employ the HLLC flux scheme \citep{Toro1994} and a five-order weighted essentially non-oscillatory slope limiter \citep{Liu1994}.
We consider optical thin radiation loss from the radiation loss curve by \cite{Colgan2008} and include thermal conduction using Spitzer conductivity. 
The saturation of thermal conduction is implemented when the electron is as fast as the sound speed \citep{Cowie1977}.
We also implement a transition-region adaptive conduction method to capture mass evaporation and energy exchange more accurately \citep{Johnston2019,Zhou2021}.
The one-dimensional hydrodynamic loop simulation based on MPI-AMRVAC has been validated in prior research \citep{Xia2011,Zhou2014,Zhou2020,Zhou2021}, which demonstrated its capability and flexibility required for this study. We note also that since flare plasma is generally confined by the magnetic field of the loops, that a 1D field-aligned approximation is a reasonable assumption \citep[for an extensive discussion of the utility of 1D flare loop models to understand the physics of flares see][]{Kerr2022}. Due to the absence of treatment of optically thick radiation, non-thermal electrons, waves, and turbulence, this code is less suitable for studies of the lower atmosphere than other (radiation-) hydrodynamics codes such as RADYN \citep{Allred2015}. Those aspects, however, are not essential to our goals and are beyond the scope of this paper which focuses on the flaring coronal dynamics.

As in most previous work, we model an individual flare loop as a semi-circular tube with a uniform cross-section.
The loop coordinate $l$ ranges from $-\pi R_\mathrm{L}/2$ to $\pi R_\mathrm{L}/2$ from one end to the other, where $R_\mathrm{L}$ is the major loop radius, and represents the loop height as well.
For the initial condition, we adopt a simplified atmospheric model with a temperature profile $T$ defined by 
\begin{equation}
T(h)=T_\mathrm{b}+\frac{T_\mathrm{t}-T_\mathrm{b}}{2}\tanh \left(\frac{h - h_\mathrm{tr}}{w_\mathrm{tr}} + 1\right),
\end{equation}
where $h$ is the height. Here $T_\mathrm{b} = 10$~kK is the temperature at the foot point, $h=0$.
This value is based on the observed chromospheric temperatures in M dwarfs, which can range from several thousand to tens of thousands of Kelvin \citep{Cram1979, Mauas2000}.
We note that RADYN radiative hydrodynamic simulations by \cite{Kowalski2016} revealed that a temperature of approximately 10,000 K is required to accurately replicate flare emission on M stars in white light.
At the loop top (apex), the loop coordinate is $l_\mathrm{t}=0$; the temperature is $T_\mathrm{t}=6$~MK, typical for M dwarf corona \citep{Allred2015}. The transition region has a height of $h_\mathrm{tr}=5$~Mm, and a thickness of $w_\mathrm{tr}=250$~km. These values are on the higher end of the typical the solar values \citep{Zhang1998,Tian2008}. Such a piecewise-like temperature profile is widely used for simplified solar atmosphere in previous studies \citep{Bradshaw2003,Fang2016,Johnston2019}. The hot coronal portion of the loop is maintained using a uniform background heating rate of $Q_\mathrm{bg}=2\times10^{-2}$~erg~cm$^{-3}$~s$^{-1}$, which is in the typical range of solar coronal heating power \citep{Sakurai2017}.

We set the number density at the loop apex as $n_\mathrm{e}=10^{10}$~cm$^{-3}$. The initial density and pressure along the loop are computed using ideal gas law assuming a hydrostatic atmosphere. The stellar parameters used here are the median of those M-dwarfs with ``peak-bump" light curves from \cite{Howard2022}: effective temperature $T_\mathrm{eff}=3323$~K, radius $R_{\star}=0.48 R_{\sun}$, and stellar surface gravity acceleration $g=5.65\times10^4$~cm~s$^{-2}$.
The model first runs with only background heating for $2\times10^4$ s, allowing the atmosphere to relax into equilibrium.

\begin{figure}[t!]
\centering
\fig{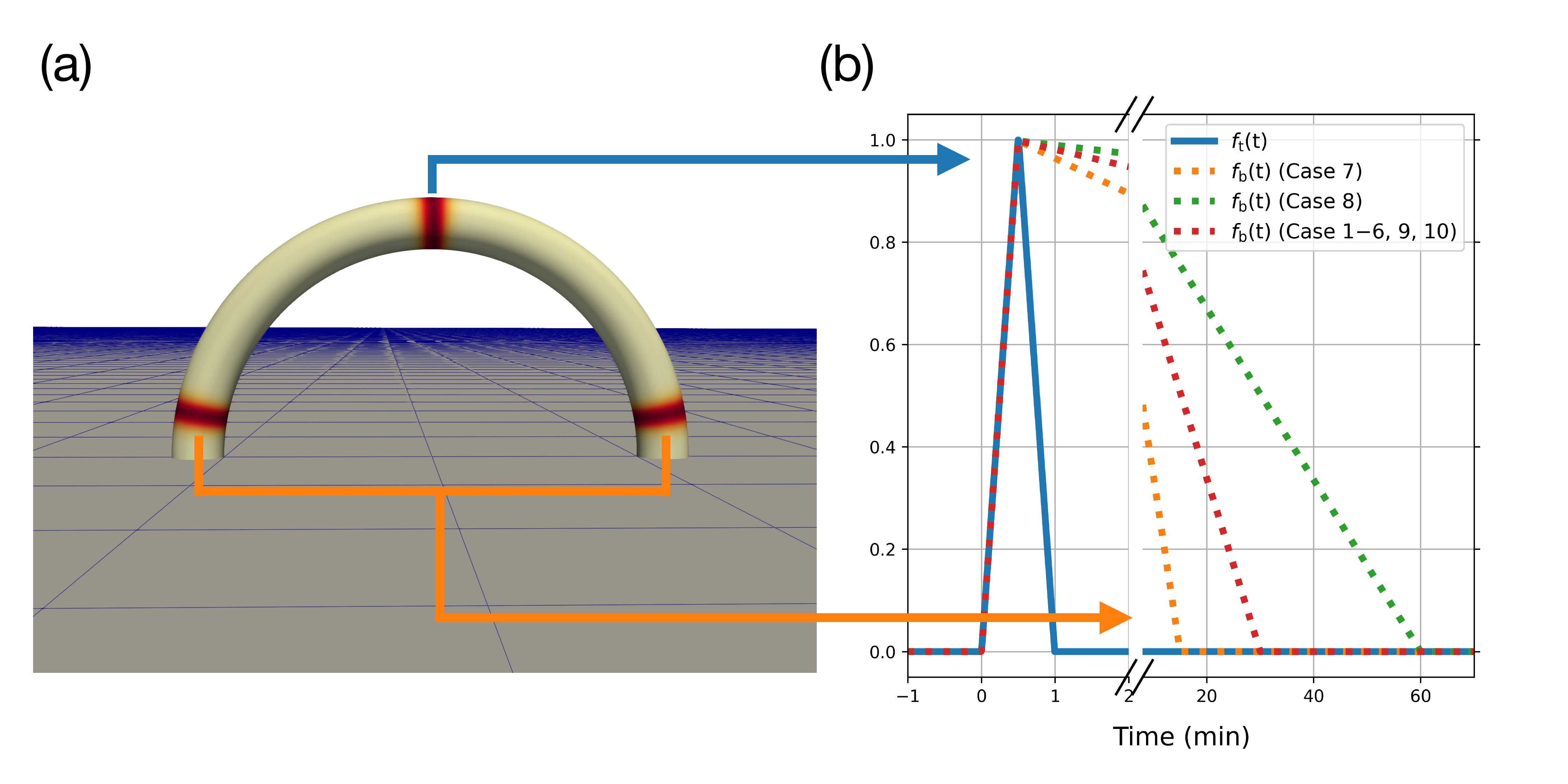}{1\textwidth}{}
\caption{(a) Illustration of our 1D loop model. The locations of the heating sources are colored in brown for the loop apex and foot points as in Eq.~(\ref{eq:heat}). (b) Normalized temporal profiles of the heating rate for the apex $f_\mathrm{t}(t)$ (blue solid line) and foot-points $f_\mathrm{b}(t)$ (dotted lines) for three different cases listed in Table~\ref{tab:1} (orange, green, red). The decay time scales of the foot-point heating rate are $14.5$, $29.5$, and $59.5$ minutes, respectively. The left and right portions depict the initial impulsive loop-top heating and the late-time gradual foot-point heating processes, respectively. They are discontinuous and have different scales in the horizontal axis.
\label{fig:2}}
\end{figure}

Following an empirical scaling law based on stellar flare observations \citep{Shibata1999,Shibata2002}, we set the nominal flare heating rate as $Q=B^2V_a/(4\pi L)$, where $B$ is the characteristic magnetic field strength, $L=\pi R_\mathrm{L}$ is the loop length, and $V_a=B/\sqrt{4\pi\rho}$ is the Alfv\'{e}n speed.
The mass density $\rho$ corresponds to a nominal electron number density $5\times 10^9$~cm$^{-3}$, which is the value at the loop top after the relaxation (c.f. $10^{10}$~cm$^{-3}$ initial value).

In this work, we investigate two loop radii (apex heights), $10$~Mm and $30$~Mm, which are typical for average and large solar active regions \citep{Benz2017} and are well consistent with stellar flare statistics \citep{Shibata1999,Shibata2002,Namekata2017}. We further consider three different magnetic field strengths, $400$, $600$, and $800$ G. The choice of $400$ G is based on the microwave observations of a coronal current sheet during a large solar flare \texttt{SOL2017-09-10T15:35} \citep{Chen2020NatAs}. The other two values are meant for the M dwarfs, where the mean photospheric field can reach kilogauss range with much larger starspots \citep{Berdyugina2005,Shulyak2019,Reiners2022}.

As visualized in Figure~\ref{fig:2}(a), we simulate the flare energy injection as time-dependent, localized heating sources at the loop apex ($H_\mathrm{t}$) and the foot-points ($H_\mathrm{b}$).
The former mimics the impulsive heating directly caused by magnetic reconnection, whereas the latter mimics a secondary, gradual heating of the lower atmosphere. The approach is widely used in 1D flare simulation (see discussion in Section~\ref{sec:discussion}).
The two heating terms are formulated as:
\begin{equation}\label{eq:heat}
\begin{split}
H_\mathrm{t}(t) &= k_\mathrm{t} \, f_\mathrm{t}(t)\,Q\,\exp\left[-(l-l_\mathrm{t})^2/\lambda^2\right], \\
H_\mathrm{b}(t) &= k_\mathrm{b} f_\mathrm{b}(t)\,Q\,\exp\left[-(h-h_\mathrm{tr})^2/\lambda^2\right]. \\
\end{split}
\end{equation}
Here $l_\mathrm{t}=0$ is the loop coordinate of the apex, and $h_\mathrm{tr}=5$~Mm is the height of the transition region as defined earlier. We use $\lambda=1$~Mm as the typical spatial scale of hydrodynamic flare heating model \citep{Bradshaw2003}, which is compatible with the calculation based on non-thermal electrons' energy deposition layer \citep{Radziszewski2020}. This value for the loop-top source is justified from the bow shock length along loops in 2D simulations \citep{Chen2015Sci,Ruan2020}. We note that the ``foot-point'' heating is applied at the transition region height, $h=h_\mathrm{tr}=5$~Mm, rather than the actual foot point $h=0$~Mm.
The total energy input $E$ is estimated as the temporally and spatially integrated heating $H_\mathrm{t}+2H_\mathrm{b}$ (note the factor of two is to account for both foot points; see notes of Table~\ref{tab:1} for more detail). The piece-wise function $f_i(t)$, where $i$ stands for loop top (t) or foot point (b), is defined as
\begin{equation}\label{eq:time}
f_i(t)=\begin{cases}
    (t-t_\mathrm{s})/\tau_\mathrm{r},& t_\mathrm{s}<t<t_\mathrm{p}\\
    1-(t-t_\mathrm{p})/\tau_{\mathrm{d},i},& t_\mathrm{p}<t<t_{\mathrm{e},i},\\
    0,              & \text{otherwise},
\end{cases}
\end{equation}
where $\tau_\mathrm{r}$ and $\tau_{\mathrm{d},i}$ indicate the time scale for the rising and decaying heating phase, respectively. Additional parameters, $t_\mathrm{s}$, $t_\mathrm{p}$, and $t_{\mathrm{e},i}$, denote the start, peak, and end times of heating, respectively. They satisfy $t_\mathrm{s}+\tau_\mathrm{r}=t_\mathrm{p}$, and $t_\mathrm{p}+\tau_{\mathrm{d},i}=t_{\mathrm{p},i}$ by definition. Finally, a scaling factor $k_i$, defined as
\begin{equation}\label{eq:scaling}
\begin{split}
    k_\mathrm{t} & = 1,\\
    k_\mathrm{b} &= \dfrac{\tau_\mathrm{r}+\tau_\mathrm{d,t}}{\tau_\mathrm{r}+\tau_\mathrm{d,b}},\\
\end{split}
\end{equation}
makes the time-integrated energy from one foot-point and the loop-top heating sources to be approximately equal.
We consider this equal-partition assumption to be appropriate given the lack of knowledge of the heating mechanism.

\begin{deluxetable*}{cccccc}[t!]
\tablenum{1}
\tablecaption{1D flare model parameters \label{tab:1}}
\tablewidth{0pt}
\tablehead{
\colhead{ } & \colhead{Loop radius} & \colhead{Magnetic field} & \colhead{Decay time\tablenotemark{a} } & \colhead{Nominal heating rate} & \colhead{Total energy\tablenotemark{b}} \\
\colhead{ } & \colhead{$R_\mathrm{L}$ [Mm]}  & \colhead{$B$ [G]} & \colhead{$\tau_\mathrm{d,b}$ [min]} & \colhead{$Q$ [erg~cm$^{-3}$~s$^{-1}$]} & \colhead{$E$ [$10^{33}$~erg]}
}
\startdata
Case 1 & $10$ & $400$ & $29.5$ & $5000$  & $1.4$  \\
Case 2 & $10$ & $600$ & $29.5$ & $16877$  & $4.7$  \\
Case 3 & $10$ & $800$ & $29.5$ & $40005$  & $11.2$ \\
Case 4 & $30$ & $400$ & $29.5$ & $1667$ &  $4.2$ \\
Case 5 & $30$ & $600$ & $29.5$ & $5626$ & $14.1$ \\
Case 6 & $30$ & $800$ & $29.5$ & $13335$ & $33.5$ \\
\hline
Case 7 & $10$ & $800$ & $14.5$ & $40005$  & $11.2$ \\
Case 8 & $10$ & $800$ & $59.5$ & $40005$  & $11.2$ \\
\hline
Case 9\tablenotemark{c} & $10$ & $800$ & $29.5$ & $40005$  & $7.5$ \\
Case 10\tablenotemark{c} & $10$& $800$ & N/A & $40005$  & $3.7$ \\
\hline
Case 11\tablenotemark{d} & $10$ & $400$ & $29.5$ & $5000$  & $1.35$  \\
Case 12\tablenotemark{d} & $10$ & $400$ & $29.5$ & $5000$  & $1.16$  \\
Case 13\tablenotemark{d} & $10$ & $400$ & $29.5$ & $5000$  & $0.98$  \\
\enddata
\tablecomments{
\tablenotetext{a}{
The decay time scale of the foot-point sources, which provides the gradual heating. In Case 10, there is no foot-point source.
}
\tablenotetext{b}{
The total energy, calculated as $\int \left[H_\mathrm{t}(t)+2H_\mathrm{b}(t)\right]\mathrm{d}t$, represents the sum of the heating at the loop top and two foot points. 
The temporal profile of the heating has a triangle shape with durations of $\tau_\mathrm{r,t}+\tau_\mathrm{d,t}$ and $\tau_\mathrm{r,b}+\tau_\mathrm{d,b}$ for $H_\mathrm{t}$ and $H_\mathrm{b}$, respectively (Eq. \ref{eq:time}).
The first term is estimated to be $0.5 \pi \lambda L^2 Q (\tau_\mathrm{r,t}+\tau_\mathrm{d,t})$ where we use $\lambda=1$~Mm as the heating scale, $L=\pi R_\mathrm{L}$ as the loop length, and $\pi L^2$ as the flaring area.
The second term is the energy from two foot points, estimated to be $k_\mathrm{b} \pi \lambda L^2 Q (\tau_\mathrm{r,b}+\tau_\mathrm{d,b})$.
The associated total energies are calculated accordingly.
}
\tablenotetext{c}{
Cases 9 and 10 only contain the foot-point and loop-top sources, respectively. 
}
\tablenotetext{d}{
Cases 11--13 contain asymmetric heat source settings. The heat sources at foot point $l=\pi R_\mathrm{L}/2$ are reduced by multiplying a factor of $0.9$, $0.5$, $0.1$, respectively.
}
}
\end{deluxetable*}

The heating time profiles used in this study are shown in Figure \ref{fig:2}(b). We assume the loop-top and the foot-point heating start and reach their peak at the same time with identical rising time scales. Specifically, we use $t_\mathrm{s}=0$~s, $t_\mathrm{p}=30$~s, and $\tau_\mathrm{r}=30$~s for both sources. For the more impulsive loop-top source, the decay time scale is fixed at $\tau_\mathrm{d,t}=30$~s. For the more gradual foot-point sources we consider three different decay time scale, $\tau_\mathrm{d,b}=14.5$, $29.5$, and $59.5$~min. The corresponding end times are $t_\mathrm{e,b}=15$, $30$, and $60$~min, respectively. We note that the $60$~s impulsive heating profile is typical for a single loop in the solar cases \citep{Bradshaw2013,Qiu2016}. The gradual heating durations, $15$ and $30$~min, are typical for the solar cases \citep{Qiu2016,Zhu2018}. The $60$~min cases are longer than what have been reported for solar flares, but may well be within the range of more energetic stellar flares.

For this study, we perform a total of ten simulations with various combinations of free parameters, as summarized in Table \ref{tab:1}. We divide these simulations into four groups. For the first group (Cases 1--6), we fix the decay time $\tau_\mathrm{d,b}=29.5$~min, and explore the effect of changing loop radius $R_\mathrm{L}$ and magnetic field $B$. For the second group (Cases 7--8), we fix $R_\mathrm{L}=10$~Mm and $B=800$~G, and explore the effect of the changing decay time scale of foot-point heating. For the third group (Cases 9--10), we experiment with having only the foot-point or loop-top heating source.
In the fourth group (Cases 11-13), we examine the impact of asymmetric foot-point heating by scaling the foot-point heating $H_b(t)$ at $l=\pi R_\mathrm{L}/2$ by factors of $0.9$, $0.5$, and $0.1$, respectively. All other parameters remain the same with Case 1.

\subsection{Emission Synthesis}\label{sec:syn}

To investigate gradual phase emission, we synthesize the optically thin continuum emission $I_\mathrm{\nu}$ assuming the loop is filled with a completely ionized, hydrogen plasma \citep{Heinzel2017,Heinzel2018,Jejcic2018}. 
We consider only contribution from the coronal portion of the loop, that is, above the critical height where $T(h)>0.2$ MK in the initial, relaxed atmosphere. We consider only the wavelength range between $534$ and $1060$~nm used by the \textit{TESS} filter.

We assume an off-limb flare loop in local thermal equilibrium and ignore the background intensity. The optically thin hydrogen emission from free-free ($I_\mathrm{\nu}^\mathrm{ff}$) and free-bound ($I_\mathrm{\nu}^\mathrm{fb}$) mechanism can be expressed as
\begin{equation}
I_\mathrm{\nu}^\mathrm{ff}+I_\mathrm{\nu}^\mathrm{bf}=B_\mathrm{\nu}(T)(\kappa_\mathrm{\nu}^\mathrm{ff}+\kappa_\mathrm{\nu}^\mathrm{bf})D,
\end{equation}
where $D$ is the geometric depth of the emitting plasma along the line of sight, assumed to be equal to the loop length $D=L$ for simplicity. Additionally, $B_\mathrm{\nu}(T)$ is the Planck function, $\kappa_\mathrm{ff}$ and $\kappa_\mathrm{bf}$ are the hydrogen free-free and bound-free opacity, given by
\begin{equation}\label{eq:kff}
\kappa_\mathrm{\nu}^\mathrm{ff}=3.69\times 10^8 \,n_\mathrm{e}^2 \, g_\mathrm{ff}(\nu,T) \, T^{-1/2} \, \nu^{-3}(1-e^{-h\nu/k_\mathrm{B}T}),
\end{equation}
\begin{equation}\label{eq:kbf}
\kappa_\mathrm{\nu}^\mathrm{bf}=1.166\times 10^{14} \, n_\mathrm{e}^2 \, g_\mathrm{bf}(i,\nu) \, i^{-3} \, \nu^{-3} \, e^{h\nu_i/k_\mathrm{B}T} T^{-3/2} (1-e^{-h\nu/k_\mathrm{B}T}).
\end{equation}
Here, $n_\mathrm{e}$ is the electron density; $i$ and $\nu_i$ represent the principle quantum number and the continuum limit frequency of the specific spectral series, respectively;
$g_\mathrm{ff}(\nu,T)$ and $g_\mathrm{bf}(i,\nu)$ are the corresponding Gaunt factors, assumed to be unity; 
and $k_B$ is the Boltzmann constant. 
The free-bound emissions are calculated from Paschen to Humphreys continua.
Furthermore, the Thomson scattering by electrons is formulated as
\begin{equation}
I_\mathrm{\nu}^\mathrm{Th}=n_\mathrm{e}\sigma_\mathrm{T}\,J_\mathrm{\nu}^\mathrm{inc}D,
\end{equation}
where $\sigma_\mathrm{T}=6.65\times 10^{-25}$~cm$^{-2}$ is the cross-section for Thomson scattering, and $J_\mathrm{\nu}^\mathrm{inc}$ is the incident intensity, which equals $B_\mathrm{\nu}(T_\mathrm{eff})$ times a dilution factor of 0.4 \citep{Heinzel2018}.
Finally, the total flare loop emission is 
\begin{equation}
I_\mathrm{\nu}=I_\mathrm{\nu}^\mathrm{ff}+I_\mathrm{\nu}^\mathrm{bf}+I_\mathrm{\nu}^\mathrm{Th}.
\end{equation}

The observed \textit{TESS} stellar flare light curve is generally normalized by the quiescent flux from the stellar disk.
To compare our model with observations, we estimate the relative flare flux by calculating the ratio between the emission from the coronal flare loop and the background emission. The latter is approximated by a uniform disk of blackbody radiation $B_\mathrm{\nu}(T_\mathrm{eff})$. As a first-order estimate, we ignore the limb-darkening effect. To determine the emission from the flare loop, we integrate $I_\mathrm{\nu}f(\nu)$ across frequency $\nu$ and along the loop $l$, multiply it by the spatial length scale $\pi L$ to emulate the effect of multiple loops. Finally, both the flare and the stellar disk flux are scaled by the \textit{TESS} filter response function $f(\nu)$ \citep{Ricker2015}:
\begin{equation}\label{eq:rflux}
    \delta F/F \approx \dfrac{L\iint I_\mathrm{\nu}\,f(\nu) \, {\rm d} \nu \, {\rm d}l}{R_{\star}^2\int B_\mathrm{\nu}(T_\mathrm{eff})\,f(\nu) \, {\rm d} \nu}.
\end{equation}

\begin{figure}[h!]
\centering
\fig{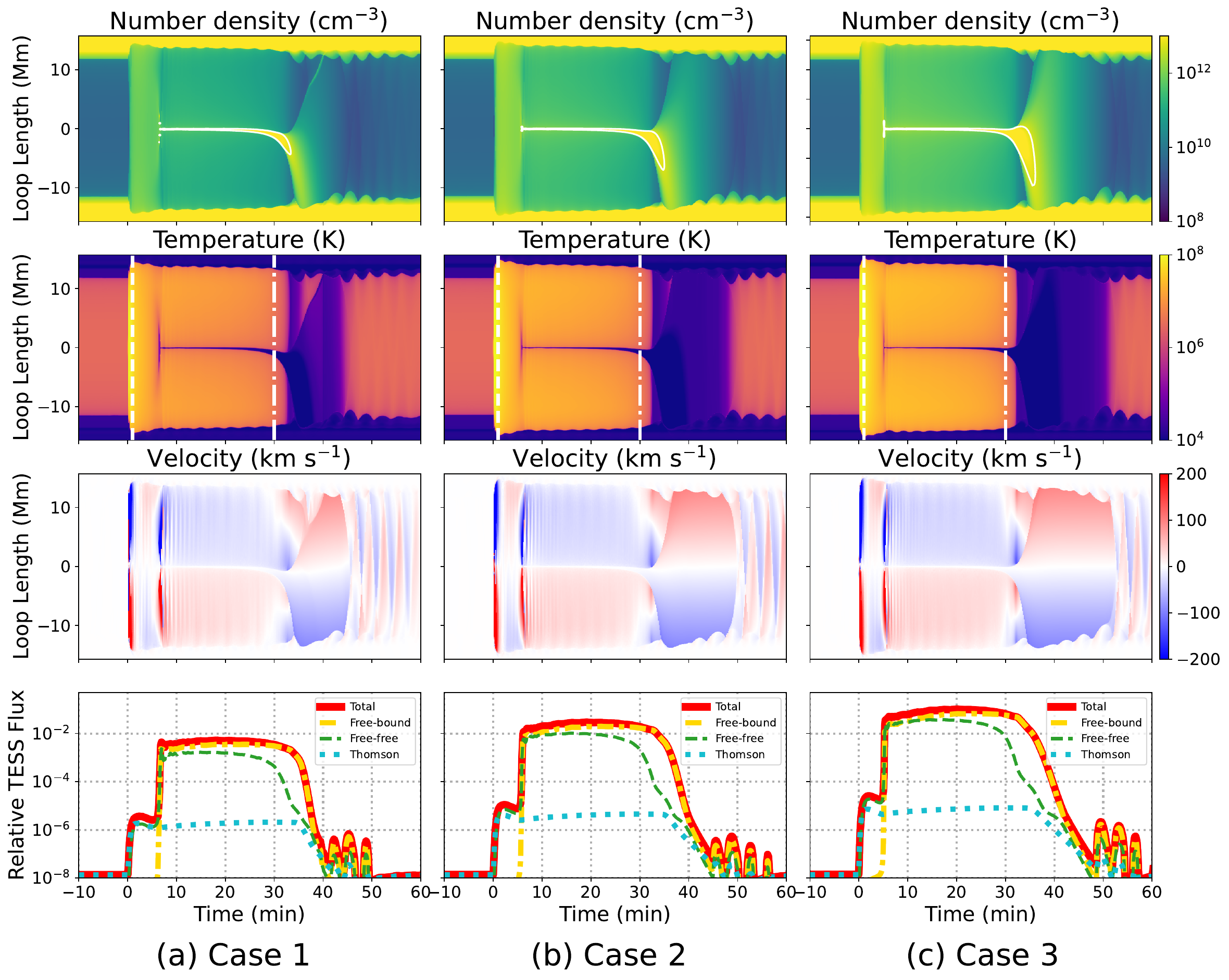}{0.95\textwidth}{}
\caption{Modeled flare atmosphere and synthesized \textit{TESS} light curves. Cases 1--3 are shown in columns (a)--(c), respectively (fixed loop radius $R_\mathrm{L}=10$~Mm, nominal foot-point heating time scale $\tau_\mathrm{d,b}=29.5$~min, and varying magnetic fields $B=400$, $600$, and $800$~G). The top three rows show the evolution of electron number density $n_\mathrm{e}$, temperature $T$, and velocity $v$ along the flare loop. The vertical axis is the loop coordinate $l$, and the horizontal axis is time $t$. The loop apex is located at the center $l=0$; the foot-points are at the two ends. Positive (negative) velocities at positive (negative) loop coordinates indicate downflows, i.e. flows from the loop-top to the foot-points, and are shown as red (blue). Time $t=0$ marks the beginning of the heating input. The contours are for high density region in the corona $n_\mathrm{e}=5\times10^{12}$~cm$^{-3}$. The dashed and dash-dot vertical lines in the second row indicate the end time of the loop-top and foot-point source, respectively. The bottom row shows the synthesized \textit{TESS} flare emission from the coronal plasma, normalized to the stellar disk emission. The total loop emission is shown in red, and the contributions from Thomson scattering, free-free emission, and free-bound emission are shown in blue, orange, and green, respectively.
\label{fig:3}}
\end{figure}

\begin{figure}[h!]
\centering
\fig{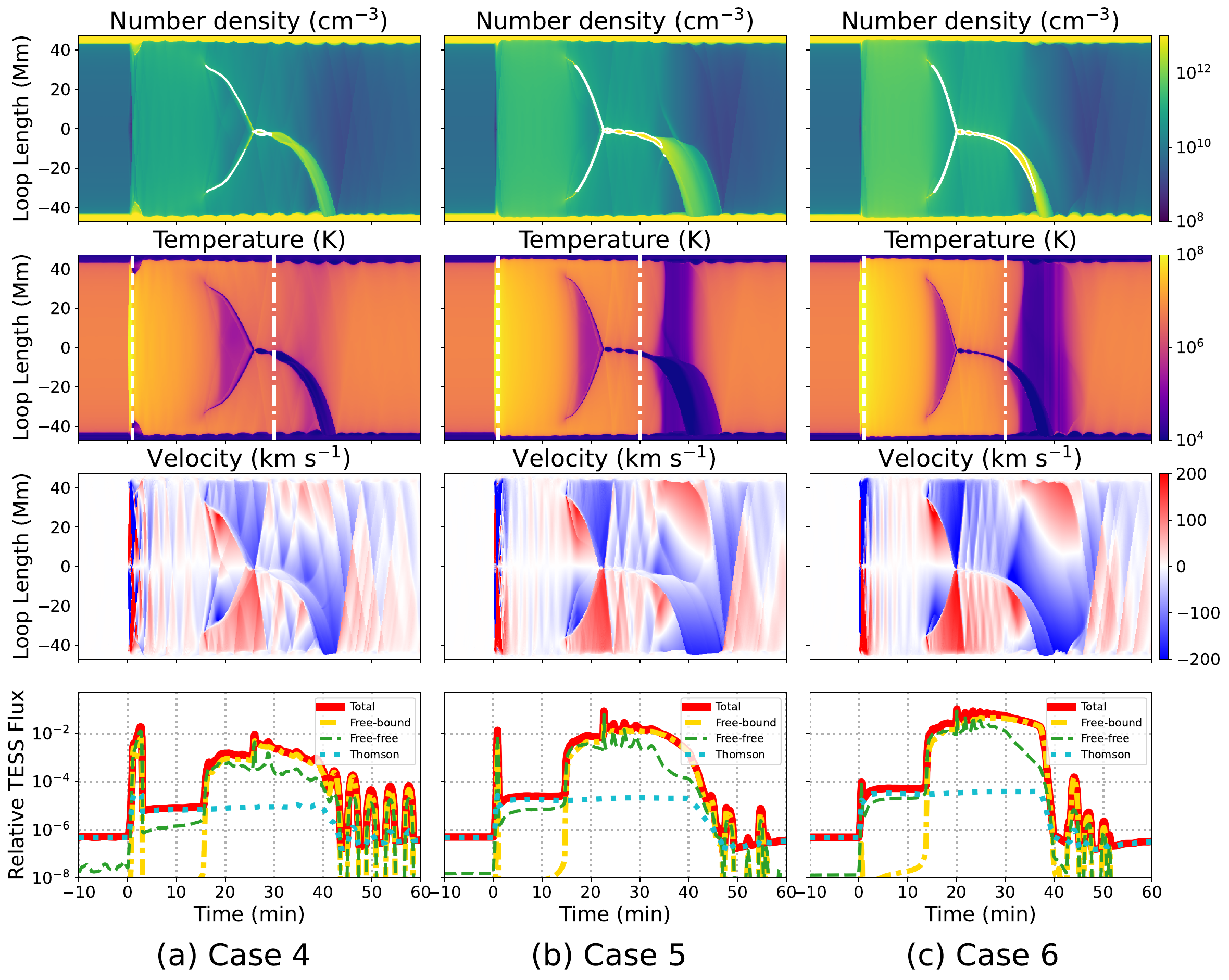}{0.95\textwidth}{}
\caption{Same as Figure~\ref{fig:3}, but for Cases 4--6 ($R_\mathrm{L}=30$~Mm).
\label{fig:4}}
\end{figure}

\begin{figure}[h!]
\centering
\fig{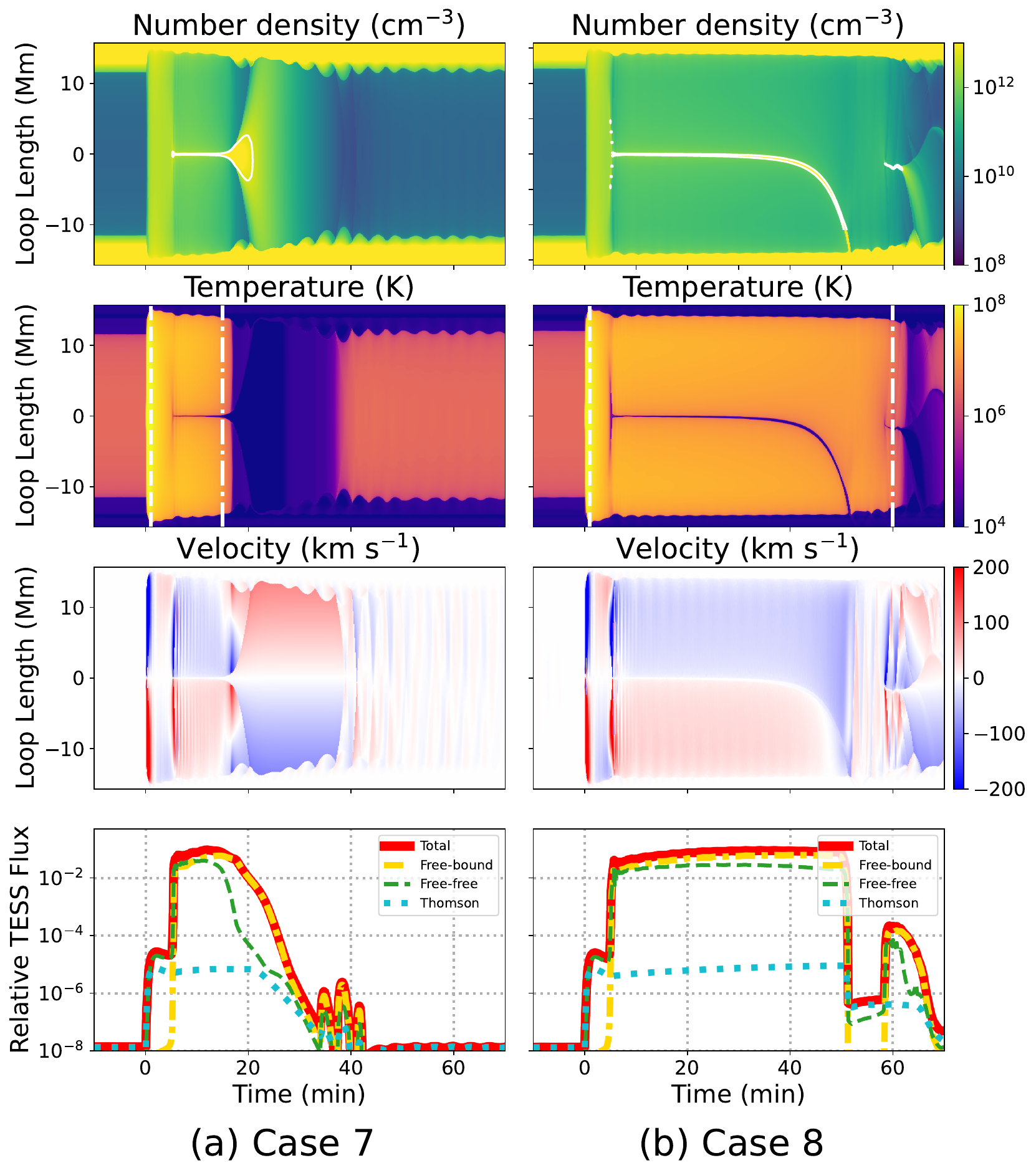}{0.8\textwidth}{}
\caption{Same as Figure~\ref{fig:3}, but for Cases 7 and 8 ($R_\mathrm{L}=10$~Mm, with half and double foot-point source heating time scales, i.e., $\tau_\mathrm{d,b}=14.5$ and $59.5$~min). They can be compared with Case 3 ($\tau_\mathrm{d,b}=29.5$~min) in Figure~\ref{fig:3}.
\label{fig:5}}
\end{figure}

\begin{figure}[h!]
\centering
\fig{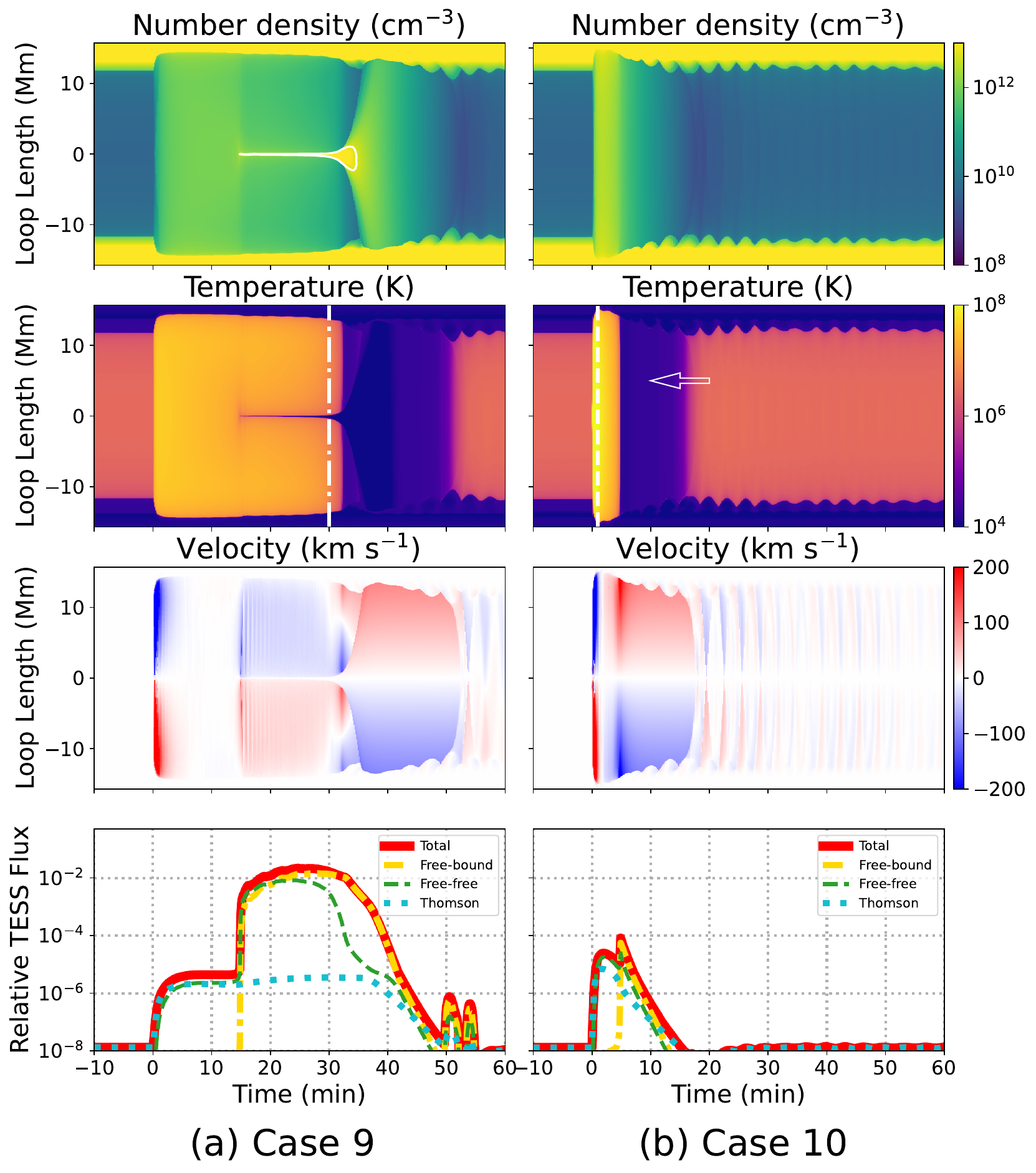}{0.8\textwidth}{}
\caption{Same as Figure~\ref{fig:3}, but for Cases 9 and 10 ($R_\mathrm{L}=10$~Mm) with only foot-point and loop-top heating sources, respectively. The white arrow points out the region with catastrophic cooling and no plasma condensation. 
\label{fig:6}}
\end{figure}

\begin{figure}[h!]
\centering
\fig{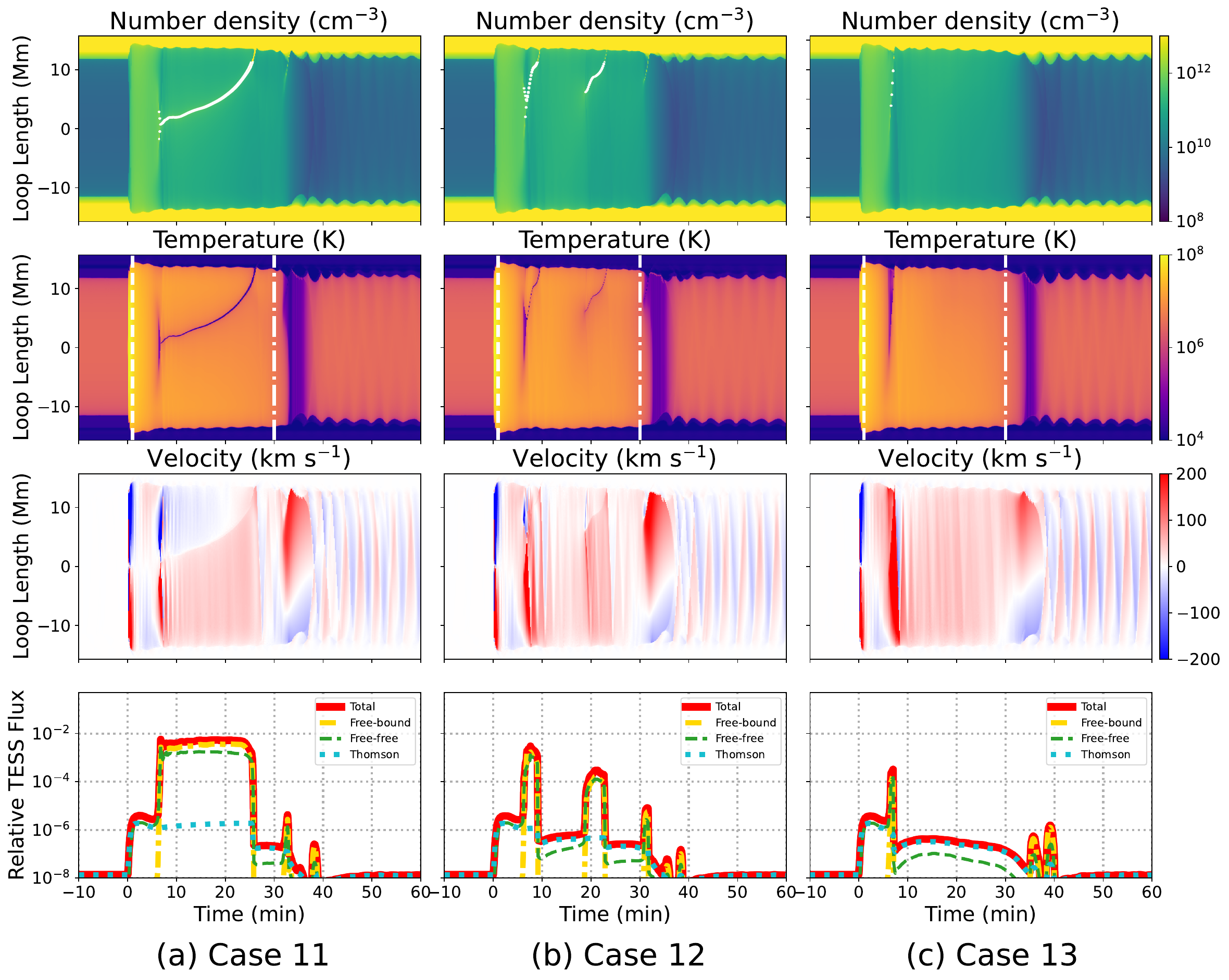}{0.95\textwidth}{}
\caption{Same as Figure~\ref{fig:3}, but for Cases 11--13 with asymmetric foot-point heat sources.
\label{fig:7}}
\end{figure}

\section{Results}\label{sec:result}

\subsection{Thermodynamic Evolution}\label{subsec:tdevo}

The modeled loop thermodynamic evolution is shown in the top three rows of Figures \ref{fig:3}--\ref{fig:7}. We use colors to visualize the evolution of the physical parameters in the loop as a function of time.
In all cases, the atmosphere reached a steady state after a relaxation period of $2\times10^4$~s before the heating source turns on at $t=0$. The two dense and cool layers near the foot points correspond to the chromosphere and transition region, while the remaining region in the central part is the hot and tenuous corona.

Shortly after the heating onset, the temperature near the loop apex $l=0$ rapidly increases to nearly $10^8$~K, whereas the density decreases briefly. This is owing to the impulsive energy injection at the loop top, and is most pronounced in Cases 4--6. In the meantime, the top of the chromosphere is heated and generates a strong evaporation flow. Shortly after, the temperature of the coronal portion of the loop rises to several tens of million Kelvin. As strong upflows toward the loop top develop due to the expansion of the chromospheric plasma, the number density of the loop gradually increases to about $10^{11}$--$10^{12}$~cm$^{-3}$ in all cases.

As the plasma cools, the increased density in the coronal loop triggers thermal instability. For shorter loops (Cases 1--3), a dramatic increase in density (to $n_\mathrm{e}>5\times10^{12}$~cm$^{-3}$) occurs near $l=0$ at $t=7$ min, which is accompanied by a rapid temperature drop to around $10^4$ K. This cool plasma remains at the loop top for an extended period of time. In contrast, cool plasma in longer loops (Cases 4--6) first appears around $l\approx\pm35$ Mm, corresponding to a height of $12$~Mm above the foot points, approximately $16$~min after the onset of heating.
The condensed plasma (white contours in the number density diagrams) is co-spatial with the front edge of the rapid upflow (same locations in the velocity diagrams).
The pattern suggests that the material is propelled by an evaporation acoustic shock with speeds ranging from $50$--$170$~km~s$^{-1}$ (c.f. local sound speed $10$--$170$~km~s$^{-1}$).

As the foot-point heating diminishes, the condensed plasma starts descending towards the chromosphere\footnote{In most cases, the plasma descends along one of the two loop legs. The asymmetry is attributed to the accumulated, but small, numerical errors.} around $t\approx 30$ min in Cases 1--6. During this phase, foot-point velocities reach approximately the free-fall value, about $100$~km~s$^{-1}$ for the shorter loops (Cases 1--3), and $200$~km~s$^{-1}$ for longer loops (Cases 4--6).
Meanwhile, global catastrophic cooling can occur elsewhere in the loop without condensation \citep{Cargill2013}. This is evidenced by the rapid decrease of temperature in Case 10 (white arrow). The temperature and density of the cool coronal plasma can also be modulated by acoustic shock waves that are being reflected in the loop. The oscillating patterns are clearly visible in the density, temperature, and velocity profiles from $20$ to $30$ min in Cases 4--6, which will result in oscillations in the synthesized \textit{TESS} light curves.

We find that the thermodynamic evolution in Cases 7 and 8 are similar to Case 3. The shorter (longer) foot-point heating duration yields an earlier (later) onset of the falling plasma. The values are about $16$ and $45$~min for Case 7 and 8, respectively. Sustained foot-point heating in Case 8 also delays global catastrophic cooling from happening in that model.

We find that the gradual foot-point heating is crucial to coronal condensation in our model, based on Cases 9 and 10.
With only foot-point heating, the evolution of Case 9 is similar to Case 3. The difference appears to be a delayed initial coronal condensation at about $15$~min.
With only a loop-top source, Case 10 does not produce any coronal plasma condensation. A catastrophic cooling phase is accompanied by a gradual decrease in density. This result is consistent with previous findings \citep{Antiochos1980,Reep2020,Antolin2022}.

Cases 11--13 demonstrate that with increased asymmetry of foot-point heating, both coronal condensation and flare emission diminish. Case 11, differing minimally from Case 1, shows a slightly reduced duration of condensed gas in the corona.  In Case 12, secondary and tertiary condensation episodes occur around the 20 and 30-minute marks, respectively. These cycles of evaporation and condensation may be attributed to thermal non-equilibrium \citep{Froment2015,Froment2017,Froment2020}. Subsequent cycles exhibit decreasing amounts of condensed plasma due to reduced foot-point heating. In the highly asymmetric Case 13, the negligible plasma condensation in the corona is accompanied by a substantial flow between foot points. This finding, characterized by a large heating ratio of 10 between the two foot points, agrees well with the results in \cite{Klimchuk2019}.

\begin{deluxetable*}{ccccccc}
\tablenum{2}
\tablecaption{Late phase features from simulations \label{tab:2}}
\tablewidth{0pt}
\tablehead{
\colhead{ } & \colhead{Time delay\tablenotemark{a}} & \colhead{Peak flux} & \colhead{Duration\tablenotemark{b}} & \colhead{Free-bound\tablenotemark{c}} & \colhead{Free-free\tablenotemark{c}} & \colhead{Thomson scattering\tablenotemark{c}}\\
\colhead{ } & \colhead{[min]} & \colhead{[$10^{-2}$ disk flux]} & \colhead{[min]} & \colhead{[\%]} & \colhead{[\%]} & \colhead{[\%]}}
\startdata
Case 1 & 18.33& 0.50 & 27.91 & 3.9 & 1.5 & 0.003\\
Case 2 & 19.18& 2.88 & 30.63 & 6.3 & 2.7 & 0.002\\
Case 3 & 20.19& 10.09 & 32.49 & 8.8 & 4.3 & 0.001\\
Case 4 & 25.91 & 0.91 & 17.32 & 0.6 & 0.3 & 0.006\\
Case 5 & 22.62& 8.04 & 25.33 & 0.9 & 0.4 & 0.004\\
Case 6 & 20.04& 10.15 & 24.19 & 1.3 & 0.6 & 0.002\\
\hline
Case 7 & 12.03& 9.02 & 18.18 & 3.4 & 1.9 & 0.0009\\
Case 8 & 34.21& 8.57 & 45.94 & 12.7 & 6.3 & 0.002\\
\hline
Case 9 & 24.62& 1.73 & 21.90 & 1.1 & 0.5 & 0.0006\\
Case 10 & N/A& N/A & N/A & 0.0003 & 0.0004 & 0.0001\\
\hline
Case 11 & 18.32& 0.52 & 18.61 & 3.1 & 1.6 & 0.002\\
Case 12 & 7.59& 0.29 & 2.15 & 0.18 & 0.14 & 0.001\\
Case 13 & N/A& N/A & N/A & 0.005 & 0.006 & 0.001\\
\enddata
\tablecomments{
\tablenotetext{a}{
Peak time of the late phase relative to the start time of flare heating.}
\tablenotetext{b}{
Duration of the late phase where magnitude is greater than $10^{-3}$ (approximate \textit{TESS} error level).}
\tablenotetext{c}{The relative energy contribution in \textit{TESS}'s optical range (corresponding to total flare energy in each case) of free-bound, free-free, and Thomson scattering emissions from the corona during the flare.}}
\end{deluxetable*}

\subsection{Signatures of Late-Phase Flare}\label{subsec:latephase}

Our synthesized light curves (Cases 1--9 and 11) qualitatively reproduce the two main features of the \textit{TESS} late-phase light curves reported in \cite{Howard2022}. First, the coronal emission from the plasma condensations exhibits a gradual peak, about $20$~minutes after the onset of impulsive heating. Second, the maximum relative fluxes of the simulated late phase range from $10^{-3}$ to $10^{-1}$. These features are quantitatively summarized in Table \ref{tab:2}.

We list below several observations of the modeled late phase flare.
\begin{itemize}
\item For a fixed loop length, greater energy injection leads to greater peak flux. For the longer loops (Cases 4--6), the peak value is reached during rapid oscillations driven by reflecting shocks.
\item Longer loop lengths lead to delayed coronal condensation and shorter late-phase duration. The peak times of the late phase do not exhibit an obvious trend.
\item Longer foot-point heating leads to longer late-phase duration.
\item The absence of gradual and/or strong asymmetric foot-point heating lead to no coronal condensation, therefore no late-phase flare emission.
\end{itemize}

We note that our models only apply to the optically thin corona. The optically thick emission from the lower atmosphere, which is believed to dominate the impulsive phase, is naturally missing from the model. The only exception is from the first few minutes in Cases 4--6 (Figure \ref{fig:4}).
The impulsive peak comes from the sudden, large amount of plasma evaporated into the lower corona ($38\lesssim|l|\lesssim42$~Mm), which is associated with the onset of heating. This is most visible in Panel (a) of Figure \ref{fig:a2} in Appendix~\ref{sec:a}.

In order to account for the optical emission from dense layers (which may very well be optically thick), one needs to properly model the flare chromosphere with sufficient spatial resolution and including non-LTE and non-equilibrium radiation transfer. That is, coupling the radiative transfer equation within the hydrodynamic equations self-consistently \citep[\eg][]{Carlsson1992,Carlsson1995,Carlsson1997,Heinzel2016IAUS..320..233H}, or employing the flux-limited diffusion approximation method \citep{Levermore1981ApJ...248..321L,Moens2022}. However, our focus is on the corona, whose thermodynamics with the optically thin radiation loss has been validated by prior research \citep{Antiochos1980,Antiochos1980ApJ...241..385A,Zhou2020}. Assessing the contribution from the denser lower atmosphere representing the impulsive phase is out of the scope of this work, which focuses the late-phase emission that we speculate originates in the corona.

In Appendix~\ref{sec:a}, we estimate the contribution of various emission mechanisms during the late phase. We find that the hydrogen free-free continuum dominates the white-light emission during the early stages, whereas the hydrogen recombination continua dominate the late stages. Thomson scattering is negligible in all cases. 

While our synthesis assumes simultaneous energization of all loops, it is important to acknowledge that the timing and relative flux of the late phase may vary if there are time delays between successive loop activations. Staggering the activations within a short time span, ranging from seconds to several minutes, could slightly extend the duration and magnitude of the late phase. This adjustment represents a slight change in the superposition of individual loops but is not expected to significantly affect the main findings of our study.

\begin{figure}[h!]
\centering
\fig{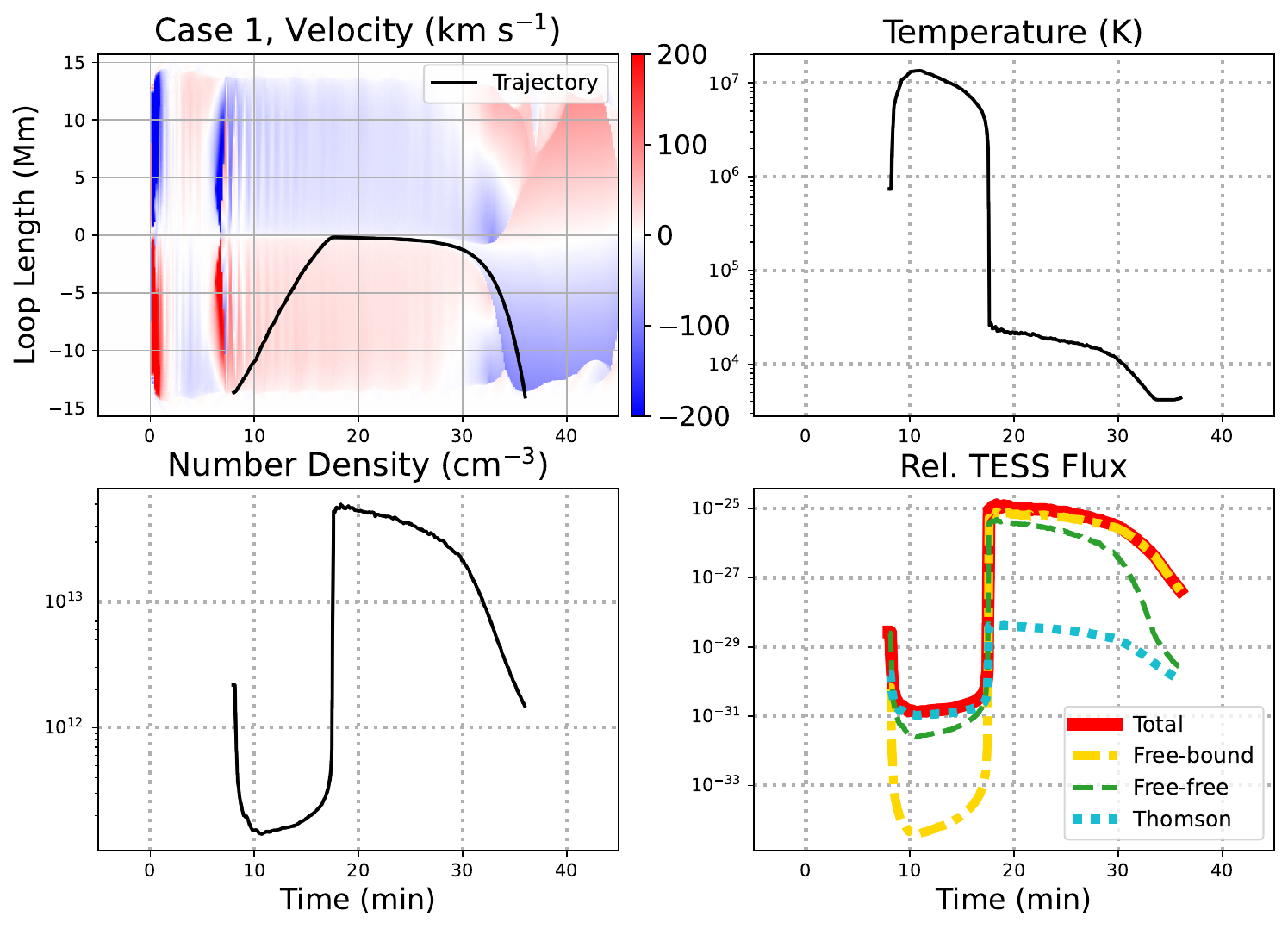}{0.95\textwidth}{}
\caption{Tracing a plasma parcel (initial position $l=-13.65$~Mm at time $t=8$~min) in Case 1 by integrating its velocity over time. Top left: trajectory of the parcel overlaid on the velocity stack plot. Top right: evolution of temperature. Bottom left: evolution number density. Bottom right: the corresponding relative \textit{TESS} flux, and contribution from free-free, free-bound, and Thomson scattering mechanisms.
\label{fig:8}}
\end{figure}

One point of interest is the life cycle of plasma and its emission signature, from its initial evaporation in the chromosphere, through condensation in the corona, to its ultimate return to the chromosphere. We investigate this by tracking a single plasma parcel in Case 1. The location of the parcel is determined by integrating the velocity in time from its initial position; its volume is simply $n_\mathrm{e}^{-1}$. The results are depicted in Figure~\ref{fig:8}.
As the parcel initially ascends (8--11~min), its density decreases sharply due to significant expansion. The temperature, however, drastically increases to above $10$~MK coronal value due to foot-point heating. The density ($2\times 10^{11}$~cm$^{-3}$) and temperature ($10^7$~K) remain relatively stable as the ascension continues. Upon reaching the loop apex (18~min), thermal instability sets in, causing the temperature to plummet from $10^7$ to $10^4$ K, and the number density to increase by two orders of magnitude. The subsequent descent is characterized by gradual temperature and density decrease.

It is evident from Figure~\ref{fig:8} that the synthetic emission correlates with the number density. This is because both free-free and bound-free opacities are proportional to the square of the number density, $\kappa_\mathrm{\nu} \propto n_\mathrm{e}^2$, as shown in Eqs.~\ref{eq:kff} and \ref{eq:kbf}. The total emission varies as $I_\nu \propto n_\mathrm{e}$ after the parcel's volume $n_\mathrm{e}^{-1}$ factors in. Furthermore, the dominating emission mechanism changes over time due to the temperature-dependent opacity for free-free and bound-free processes, which vary as $T^{-1/2}$ and $T^{-3/2}e^{h\nu_i/k_\mathrm{B}T}$, respectively (ignoring the temperature-dependence in Gaunt factors and $1-e^{-h\nu/k_\mathrm{B}T}$). After the thermal instability occurs, the temperature decrease results in a reduction of overall emission. The free-bound emission declines less significantly than the free-free emission, so the it dominates the late phase, especially between $30$ and $35$ minutes. In fact, we show that the late-phase emission predominantly stems from the cold, condensed coronal plasma cooler than $0.158$~MK, with contributions from the hotter plasma being negligible (Appendix~\ref{sec:a} and Figure~\ref{fig:a5}).

\section{Summary and Discussion}\label{sec:discussion}
In this paper, we perform a suite of 1D HD simulations for M-dwarf flares with energy from about $10^{33}$ to $10^{34}$ ergs. We assess the optically thin emission from the coronal plasma \citep{Heinzel2017,Heinzel2018}, and find that coronal plasma condensation can lead to significant emission in optical continuum. The synthetic light curves exhibit a pronounced secondary peak, whose delay time (from the initial flare heating) and the relative magnitude are qualitatively consistent with the observations from \cite{Howard2022}. We thus propose coronal plasma condensation as a possible mechanism for the late phase observed in \textit{TESS} flare light curves.

Our simulations suggest that the late phase magnitude increases with flare energy. A substantial amount of heating is required to bring the chromospheric plasma into the corona before condensation can take place, and to raise the electron density sufficiently high to produce meaningful white light emission. The least energetic event (Case 11) here has a late phase of merely $0.5\%$ of the stellar flux, but requires $1.35\times10^{33}$~erg of energy input, about $10$ times more energetic than the most intense solar flares such as \texttt{SOL2017-09-06T12:00}. This may explain why the white-light coronal loops are rarely observed in solar eruptions. Furthermore, the Sun is brighter than the M dwarf we studied, and the lack of systematic observation of these factors might also contribute to the rare sightings of these phenomena.

Our findings, as outlined in Table \ref{tab:2}, demonstrate that the energy emitted through the free-bound mechanism within \textit{TESS}'s optical range is about twice as large as that emitted through the free-free mechanism. Notably, the contribution of Thomson scattering to the total flare energy is minimal. On the other hand, the coronal plasma condensation process gives rise to a substantial amount of compressed cold plasma, predominantly situated at the loop's apex, creating a density that is two to three orders denser than the hot counterpart, as demonstrated in Figures \ref{fig:3}--\ref{fig:7}. Under the theoretical framework provided by \cite{Heinzel2018} and \cite{Jejcic2018}, the collision rate is shown to be proportional to the number density, resulting in the free-free and free-bound emission being directly proportional to the square of the density (see Eq. \ref{eq:kff} and \ref{eq:kbf}). The dependence of emission on temperature, on the other hand, is moderate. This is evidenced by a maximum tripling in emission when the temperature declines from 1 to 0.01 million Kelvin (refer to Figures 4 and 5 in \citealt{Jejcic2018}). We can intuitively understand the importance from the cold plasma through the following simple calculation. Assuming the condensed plasma accounts for a mere $1\%$ volume at the total loop, owing to its high density, the aggregate emission from these condensed materials would be between 100 to 10000 times greater than that from the rest of the loop's hot portion, a difference starkly portrayed in Figure \ref{fig:a4}.

We note that the emission synthesis is carried out separately from the simulation, making it not fully self-consistent with thermal instability process in the simulation. Detailed exploration of the consequences is beyond our the scope of this work. Still, it is important to acknowledge the critical role played by the condensed cold plasma component in generating the observed \textit{TESS} flux throughout the entire late phase. 

The simulated late-phase flare is accompanied by fast plasma evaporation and draining, whose velocities range from $50$ to $150$~km~s$^{-1}$. These motions may be probed with optical and UV spectral lines that are sensitive to relevant temperatures. Fast Doppler velocities have been reported for stellar flare observations, from several tens to hundreds km s$^{-1}$, that might have resulted from the chromospheric evaporation and/or coronal rain \citep{Argiroffi2019,Wu2022,Namizaki2023}.

The reflecting acoustic shocks in the modeled flare loops modulates the plasma density and temperature, causing the synthetic light curves to exhibit oscillations reminiscent of those in \textit{TESS} flare observations. The oscillation periods in Case 4--6 are approximately $1.5$--$2.5$ minutes, which falls within the observed range \citep[$2$--$36$~minutes;][]{Howard2022}. This may serve as a possible mechanism for the QPPs observed in some stellar flares \citep{Nakariakov2009,VanDoorsselaere2016,Zimovets2021}. Our simplified model of course cannot account for the complex MHD wave modes that are known to be important to flare dynamics.

In this study, we use loop-top and foot-point heating sources to mimic the flare energy injection. Physically, the loop-top heating is directly related to the outflow of the magnetic field reconnection, either from the bow shock on the loop-top or the collision of energetic particles from the reconnection \citep{Masuda1994,Shibata1995,Forbes1996,Guidoni2010,Fleishman2022,2020ApJ...894..148U,2021ApJ...923..248U}. The foot-point heating, on the other hand, can be attributed to turbulence dissipation of Alfv\'enic waves following loop retraction \citep[e.g.][]{Ashfield2023}, the thermalisation of energetic particles, that is nonthermal electrons or protons, \citep[e.g.][]{1971SoPh...18..489B,1978ApJ...224..241E,2011SSRv..159..107H}, Alfv\'en wave dissipation \citep[e.g.][]{1982SoPh...80...99E,2008ApJ...675.1645F,2016ApJ...827..145R,Kerr2016}, thermal conduction, and other waves-particle processes \citep{Kowalski2023}. Many recent flare models such as \texttt{RADYN} \citep{Carlsson1992,Carlsson1995,Carlsson1997,Allred2005,Allred2015} and \texttt{HYDRAD} \citep{Reep2013}, typically adopt non-thermal particles as the energy injection mechanism (though they have explored alternatives). Thus, they tend to focus on the initial impulsive footpoint heating, and it has been recognised that they do not capture the longer-duration gradual phase \citep[see discussions in][]{Allred2022,Kerr2022,Kerr2023,Reep2020}. Investigating the physical nature of these heating sources is beyond the scope of this paper. Instead, we take a mechanism-agnostic approach with our experiments, being concerned primarily with the magnitude of heating and the resulting effects.

The extended, gradual foot-point heating appears to be crucial for triggering coronal condensation\footnote{Recently, coronal condensations were self-consistently reproduced in multi-dimensional MHD simulations without artificial foot-point heating \citep{Cheung2019, Ruan2021, Chen2022}. The requirement of sustained, gradual heating could be a limitation of the symmetric 1D HD simulation.}. \cite{Reep_2020} found that foot-point heating by electron beams alone cannot produce coronal condensations in flare simulations. Some other mechanism must act alongside impulsive foot-point heating, further motivating our deposition of energy directly into the footpoint portion of our flare loop that continues energy transport through the gradual phase.

Evidence supporting the gradual heating phase is abundant in solar observations \citep{Qiu2016,Zhu2018}. It could be due to wave turbulence \citep{Ashfield2023}, turbulent suppression of thermal conduction \citep{Allred2022}, or long-lasting magnetic reconnection \citep{Ruan2021}.
For long-lasting magnetic reconnection, an extended current sheet trailing a coronal mass ejection may be required \citep{Chen2020NatAs}.

For the foot-point sources, pronounced heating asymmetry diminishes the likelihood of thermal non-equilibrium, as evidenced by our Cases 11--13 and \cite{Klimchuk2019}. As the energy input is not expected to be symmetric in most flare loops, this could account for the infrequent occurrence of late-phase activity in many super flares.

It is worth noting that we do not consider the variation of the foot-point sources location, which has been reported in solar flares \citep{Radziszewski2020}. Our justifications are threefold. Firstly, these variations are due to non-thermal electrons propagating from the corona to the lower atmosphere, losing energy in the upper and middle chromosphere via thick-target collisions. Deposition of energy in deeper layers is also possible via the beam's interaction with Langmuir and ion-acoustic waves \citep{Kowalski2023}. However, the electron beam alone does not appear to be sufficient to induce coronal condensation in HD flare modeling \citep{Reep_2020}. Secondly, while there is strong evidence of non-thermal electrons during the impulsive phase in solar flares, our focus here is rather the gradual phase. There is no compelling evidence for substantial amounts of non-thermal electrons in these later stages. Thirdly, since the time variation for foot-point source locations is not well constrained, we opt to use a constant height as in this exploratory study. The location at $h_\mathrm{tr}=5$ Mm and the thickness $\lambda=1$ Mm aligns with the height from turbulence heating in \cite{Ashfield2023} and the thickness in \cite{Radziszewski2020}.

Our emission synthesis assumes local thermal equilibrium condition and optically thin continuum radiation, which can be valid in a large range of coronal plasma parameters \citep{Heinzel2017,Heinzel2018}. It lacks proper treatment of the dense lower layers, which has been touched upon by prior studies of impulsive chromospheric flare sources using radiation hydrodynamic simulations \citep[e.g.][]{2016ApJ...816...88K,2015SoPh..290.3487K,2017ApJ...836...12K,2022FrASS...934458K}. We therefore do not model the impulsive phase of the flare, and focus on the late phase in the corona instead. We find that the maximum optical depth of the condensed plasma is around unity\footnote{We assume that all the off-limb flare loops are co-aligned along the line-of-sight, and are synchronously evolving. This leads to an estimate of the optical depth $\tau_{\nu}=\kappa_{\nu}D$.}.

In reality, the condensed plasma in the flare loops will not be perfectly aligned along the line of sight; they will also not evolve synchronously. The total emission will then come from a superposition of many segments of optically thin plasma: our conclusions are thus expected to hold qualitatively.


To further improve our understanding of stellar flares, additional observations are needed. Combining spectroscopic observations with the \textit{TESS} white-light data could provide valuable insights. Spectroscopy can offer detailed information about the temperature, density, and composition of the flaring plasma, complementing broad-band photometric observations. Moreover, EUV observations could help determine if the EUV late phase observed in solar flares is also present in stellar flares. By studying these aspects in greater detail, we can gain a more comprehensive understanding of the physical processes underlying stellar flares.

The authors are grateful to the anonymous referee for the
help comments. We thank Dana Longcope, Jeffrey Reep, and Yuhao Zhou for helpful discussions. K. E. Yang and X. Sun are supported by NASA HSOC award \#80NSSC20K1283, NSF CAREER award \#1848250, and the state of Hawai`i. GSK acknowledges the financial support from a NASA Early Career Investigator Program award \#80NSSC21K0460. The technical support and advanced computing resources from University of Hawaii Information Technology Services - Cyberinfrastructure, founded in part by the National Science Foundation CC$^*$ awards \#2201428 and \#2232862 are gratefully acknowledged.

\clearpage

\appendix

\section{Contribution of Various Emission Mechanisms}\label{sec:a}

\begin{figure}[h!]
\centering
\fig{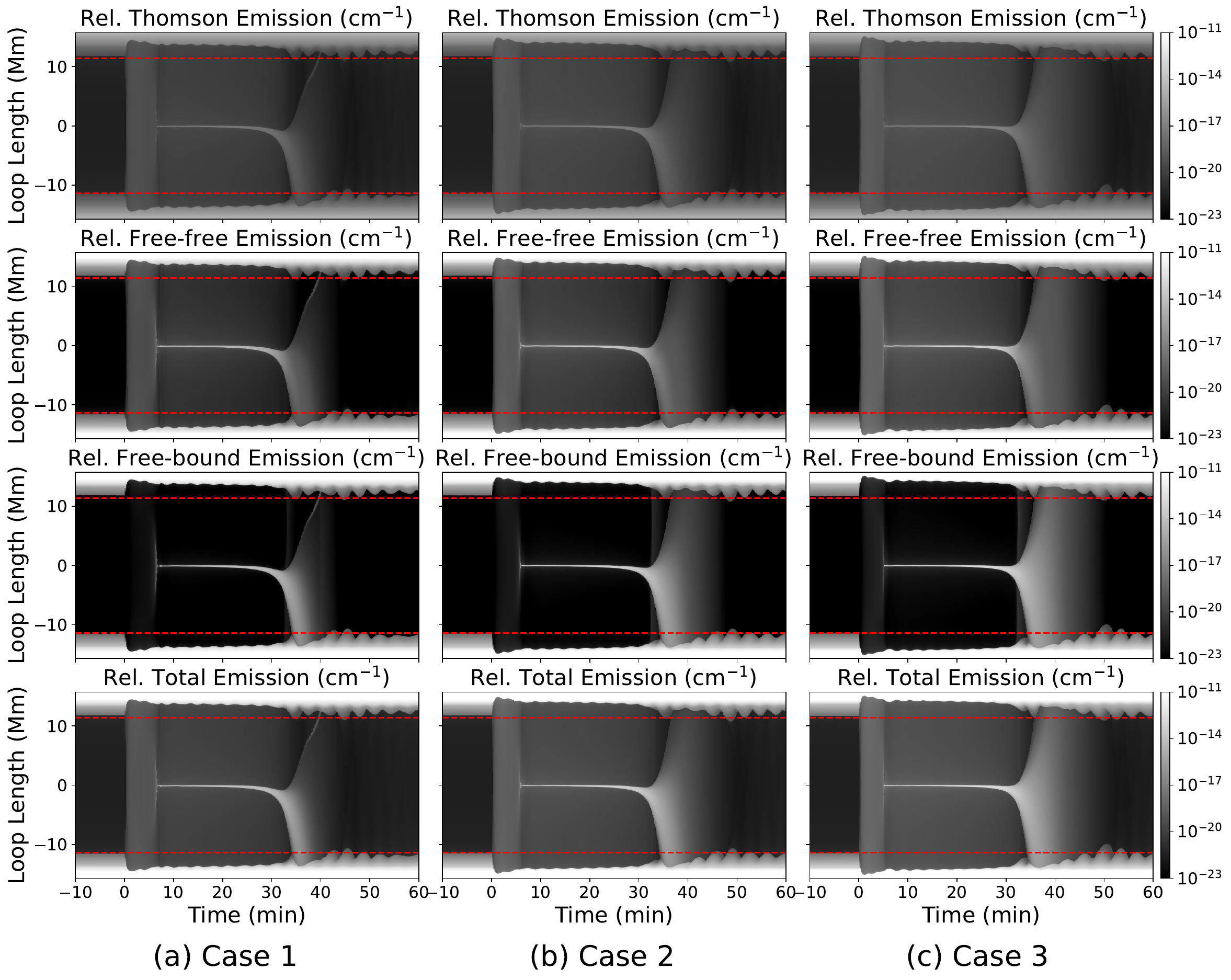}{\textwidth}{}
\caption{The relative emission contribution along the loop and its evolution in time, which is evaluated from Eqs.~\ref{eq:rflux_dis1}, \ref{eq:rflux_dis2}, and \ref{eq:rflux_dis3}. The area between the two red dashed lines indicates the corona portion.
\label{fig:a1}}
\end{figure}

\begin{figure}[h!]
\centering
\fig{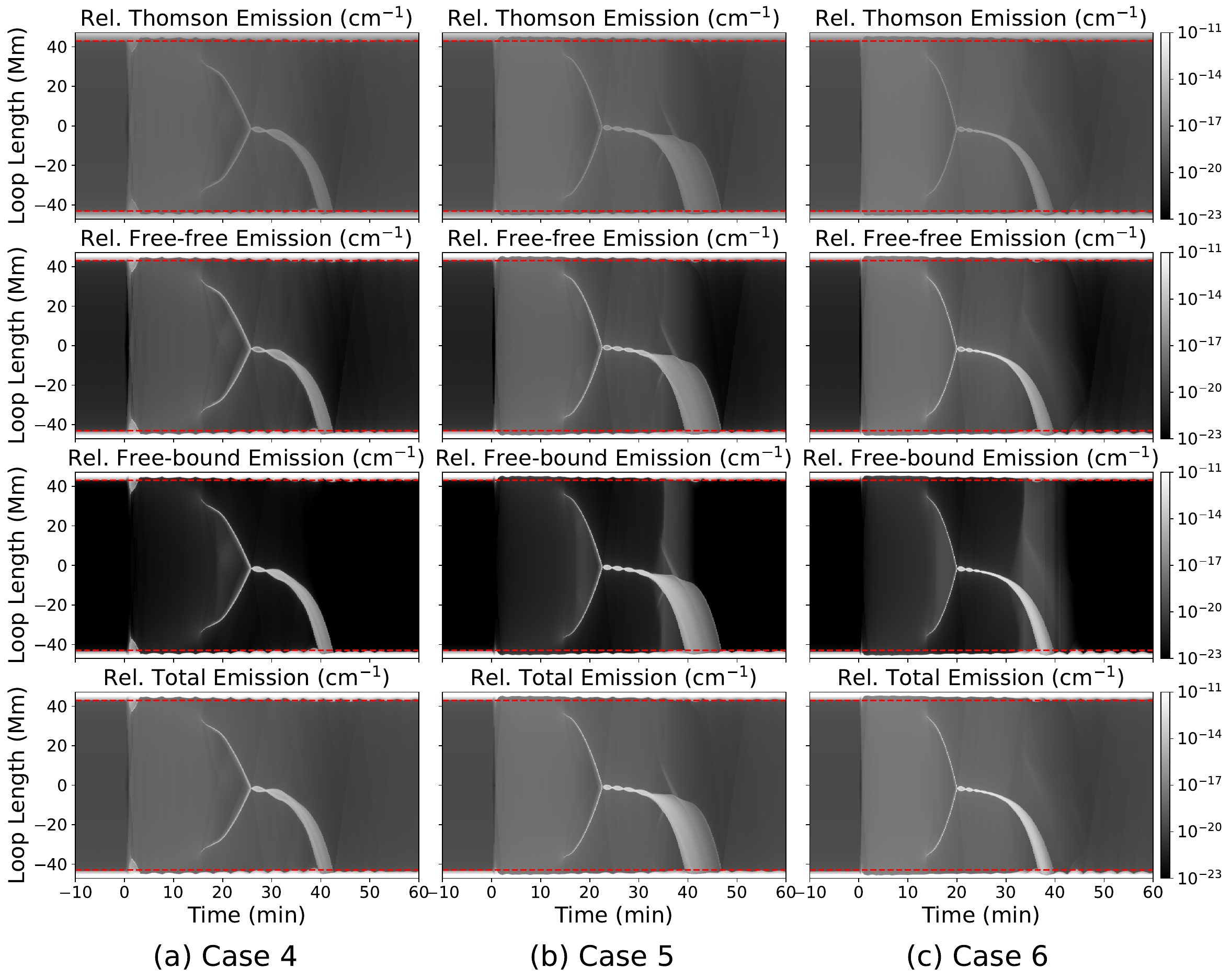}{\textwidth}{}
\caption{Same as Figure \ref{fig:a1} but for Cases 4--6.
\label{fig:a2}}
\end{figure}

\begin{figure}[h!]
\centering
\fig{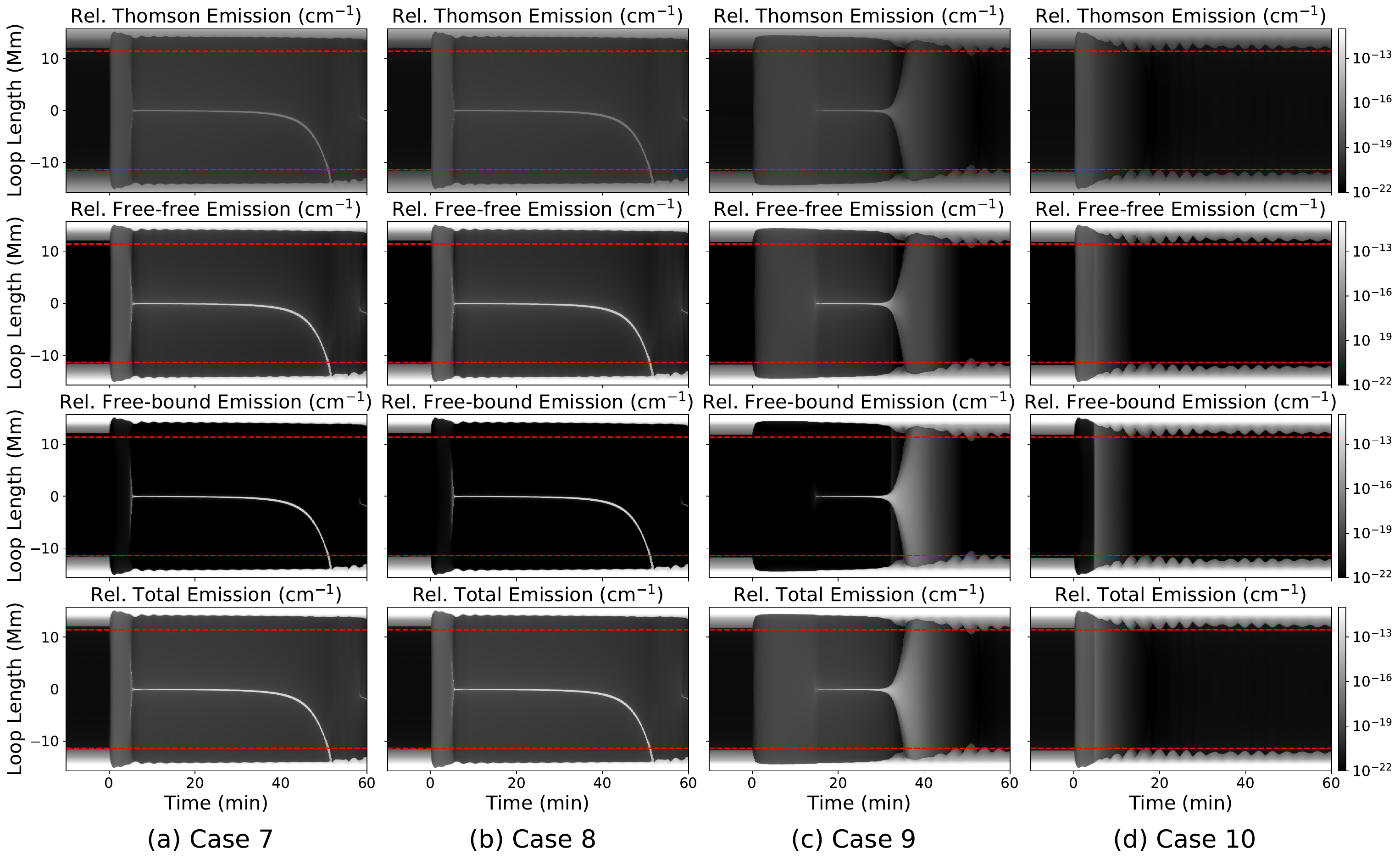}{\textwidth}{}
\caption{Same as Figure \ref{fig:a1} but for Cases 7--10.
\label{fig:a3}}
\end{figure}

\begin{figure}[h!]
\centering
\fig{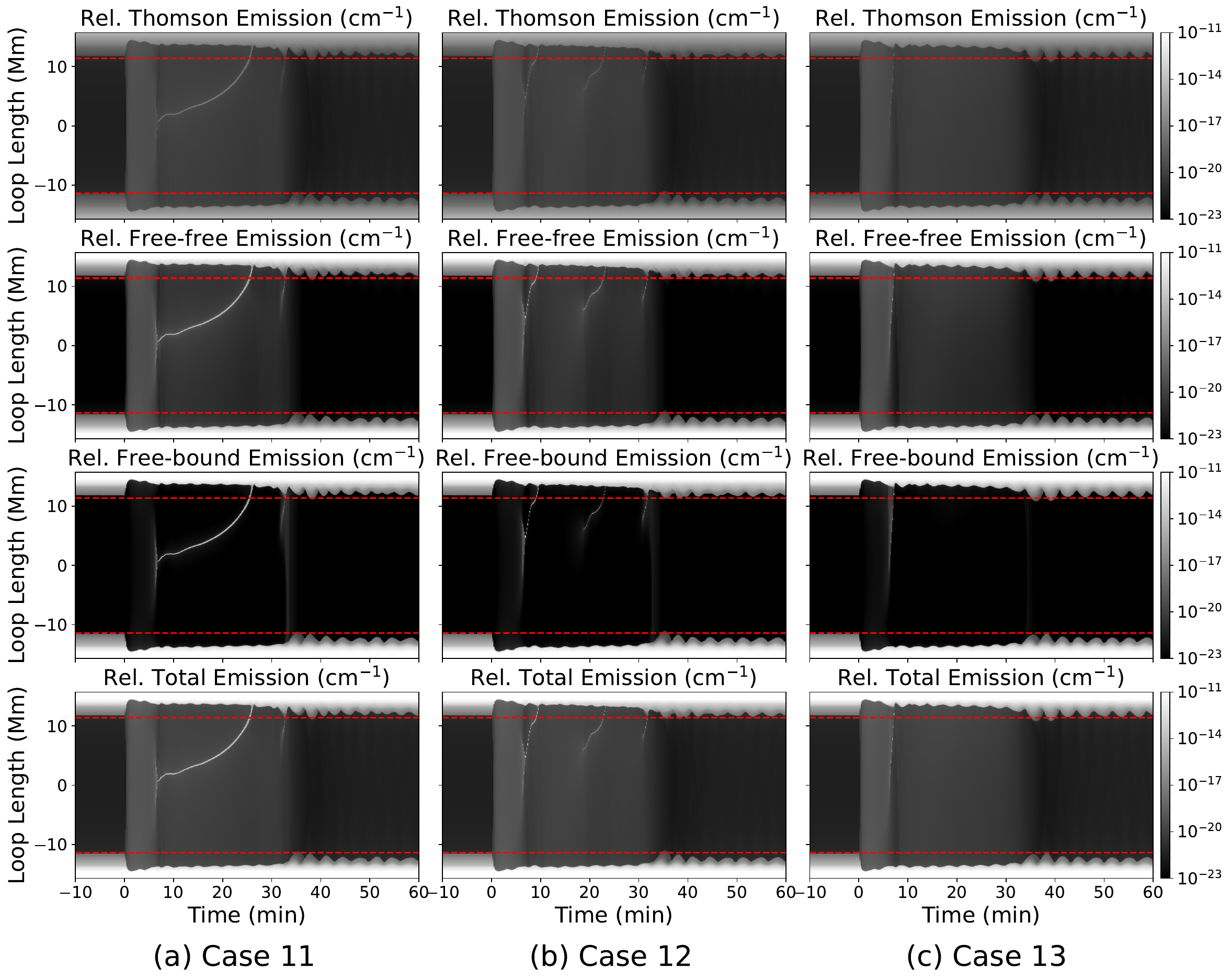}{\textwidth}{}
\caption{Same as Figure \ref{fig:a1} but for Cases 11--13.
\label{fig:a4}}
\end{figure}

The relative contribution from various emission mechanisms along the flaring loop are calculated as:
\begin{equation}\label{eq:rflux_dis1}
    \delta F^\mathrm{Th}/F \approx \frac{L \int I_{\nu}^\mathrm{Th}\,f(\nu){\rm d} \nu}{R_{\star}^2\int B_\mathrm{\nu}(T_\mathrm{eff})\,f(\nu){\rm d} \nu},
\end{equation}
\begin{equation}\label{eq:rflux_dis2}
    \delta F^\mathrm{bf}/F \approx \frac{L \int I_{\nu}^\mathrm{bf}\,f(\nu){\rm d} \nu}{R_{\star}^2\int B_\mathrm{\nu}(T_\mathrm{eff})\,f(\nu){\rm d} \nu},
\end{equation}
\begin{equation}\label{eq:rflux_dis3}
    \delta F^\mathrm{ff}/F \approx \frac{L \int I_{\nu}^\mathrm{ff}\,f(\nu){\rm d} \nu}{R_{\star}^2\int B_\mathrm{\nu}(T_\mathrm{eff})\,f(\nu){\rm d} \nu}.
\end{equation}
The total relative emission $\delta F/F$ is defined as $\delta F^\mathrm{Th}/F + \delta F^\mathrm{bf}/F +\delta F^\mathrm{ff}/F$.
Figures~\ref{fig:a1}, \ref{fig:a2}, \ref{fig:a3}, and \ref{fig:a4} show the relative emission evolution defined in the above Eq. \ref{eq:rflux_dis1}, \ref{eq:rflux_dis2}, and \ref{eq:rflux_dis3} along the flare loop for all cases.
The synthesized \textit{TESS} light curves in Figures~\ref{fig:3}, \ref{fig:4}, and \ref{fig:5} are computed by integrating the aforementioned emissions within the corona region. The corona region is defined as the area between the two red dashed lines, which indicate the location where the temperature reaches $1$~MK prior to the onset of heating.

To differentiate the contribution from the hot and cold materials in the corona, we apply a temperature threshold of $0.158$ MK, equivalent to the hydrogen ionization energy of $13.6$ eV. The emission calculations are limited to the corona region displayed in Figure \ref{fig:a1}, as illustrated in Figure \ref{fig:a4}.

\begin{figure}[h!]
\centering
\fig{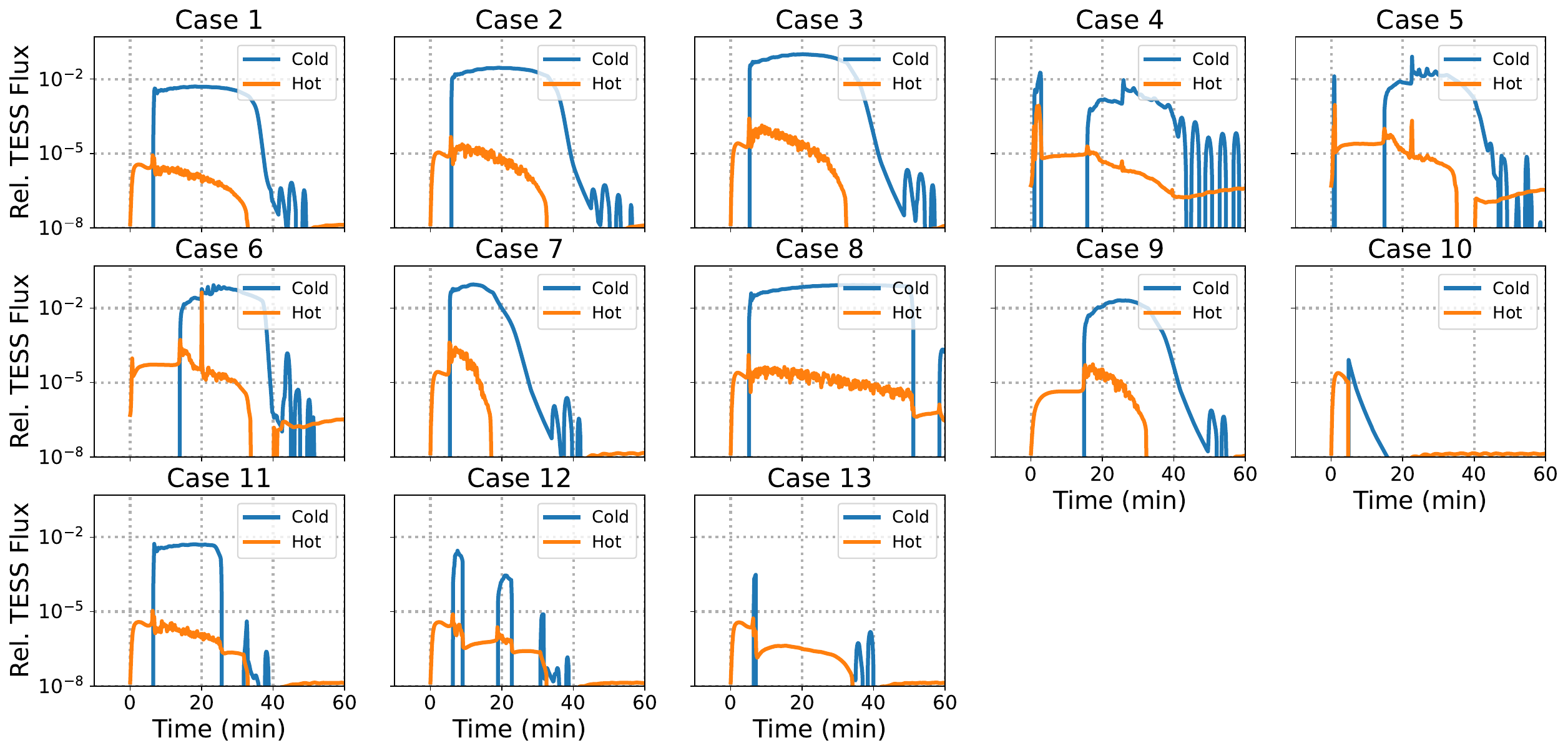}{\textwidth}{}
\caption{The emission contribution from the hot and cold materials (above or below $0.158$~MK) in the corona for all cases. \label{fig:a5}}
\end{figure}

\clearpage

\bibliography{sample631}{}
\bibliographystyle{aasjournal}



\end{CJK*}

\end{document}